\documentclass[twocolumn]{aastex631}

\NewPageAfterKeywords

\definecolor{ls4Brown}{HTML}{A0522D}

\usepackage{xparse}
\usepackage{xcolor}
\usepackage{subfigure}
\usepackage[utf8]{inputenc}

\definecolor{Maroon}{HTML}{800000}

\newcommand{\fov}{$\sim$20\,$\deg^{2}$}

\begin{document}

\title{The La Silla Schmidt Southern Survey}

\correspondingauthor{A.~A.~Miller}
\email{amiller@northwestern.edu}


\author[0000-0001-9515-478X]{Adam~A.~Miller}
\affiliation{Department of Physics and Astronomy, Northwestern University, 2145 Sheridan Rd, Evanston, IL 60208, USA}
\affiliation{Center for Interdisciplinary Exploration and Research in Astrophysics (CIERA), Northwestern University, 1800 Sherman Ave, Evanston, IL 60201, USA}
\affiliation{NSF-Simons AI Institute for the Sky (SkAI), 172 E. Chestnut St., Chicago, IL 60611, USA}

\author[0000-0002-0287-3783]{Natasha~S.~Abrams}  \affiliation{Department of Astronomy, University of California, Berkeley, CA 94720-3411, USA}

\author{Greg~Aldering}\affiliation{Lawrence Berkeley National Laboratory, 1 Cyclotron Road, MS 50B-4206, Berkeley, CA 94720, USA}

\author[0000-0003-3768-7515]{Shreya Anand} \altaffiliation{LSST-DA Catalyst Postdoctoral Fellow} \affiliation{Kavli Institute for Particle Astrophysics and Cosmology, Stanford University, 452 Lomita Mall, Stanford, CA 94305, USA} \affiliation{Department of Astronomy, University of California, Berkeley, CA 94720-3411, USA}

\author[0000-0002-4269-7999]{Charlotte~R.~Angus} \affiliation{Astrophysics Research Centre, School of Mathematics and Physics, Queens University Belfast, Belfast BT7 1NN, UK}

\author[0000-0001-7090-4898]{Iair~Arcavi} \affiliation{School of Physics and Astronomy, Tel Aviv University, Tel Aviv 69978, Israel}

\author[0000-0003-0424-8719]{Charles~Baltay}\affiliation{Department of Astronomy, Yale University, P.O. Box 208101, New Haven, CT 06520-8101, USA}

\author[0000-0002-8686-8737]{Franz~E.~Bauer} \affiliation{Instituto de Alta Investigaci{\'{o}}n, Universidad de Tarapac{\'{a}}, Casilla 7D, Arica, Chile}

\author[0000-0001-6415-0903]{Daniel Brethauer} \affiliation{Department of Astronomy, University of California, Berkeley, CA 94720-3411, USA}

\author[0000-0002-7777-216X]{Joshua~S.~Bloom} \affiliation{Department of Astronomy, University of California, Berkeley, CA 94720-3411, USA}\affiliation{Lawrence Berkeley National Laboratory, 1 Cyclotron Road, MS 50B-4206, Berkeley, CA 94720, USA}

\author[0009-0007-4271-6444]{Hemanth~Bommireddy} \affiliation{Department of Astronomy, Universidad de Chile, Camino el Observatorio 1515, Las Condes, Santiago, Chile}

\author[0000-0001-6003-8877]{Márcio Catelan} \affiliation{Instituto de Astrofísica, Pontificia Universidad Católica de Chile, Av. Vicuña Mackenna 4860, 7820436 Macul, Santiago, Chile} \affiliation{Millennium Institute of Astrophysics (MAS), Nuncio Monse\~nor S\'otero Sanz 100, Providencia, Santiago, Chile} \affiliation{Centro de Astro-Ingeniería, Pontificia Universidad Católica de Chile, Av. Vicuña Mackenna 4860, 7820436 Macul, Santiago, Chile}

\author[0000-0002-7706-5668]{Ryan~Chornock} \affiliation{Department of Astronomy, University of California, Berkeley, CA 94720-3411, USA}

\author[0000-0002-6576-7400]{Peter~Clark} \affiliation{School of Physics and Astronomy, University of Southampton, Southampton, SO17 1BJ, UK}

\author[0000-0001-5564-3140]{Thomas~E.~Collett}\affiliation{Institute of Cosmology and Gravitation, University of Portsmouth, Burnaby Road, Portsmouth, PO1 3FX, UK}

\author[0000-0001-9494-179X]{Georgios~Dimitriadis} \affiliation{Department of Physics, Lancaster University, Lancaster LA1 4YB, UK}

\author[0009-0007-8485-1281]{Sara Faris} \affiliation{School of Physics and Astronomy, Tel Aviv University, Tel Aviv 69978, Israel}

\author[0000-0003-3459-2270]{Francisco F\"orster}\affiliation{Data and Artificial Intelligence Initiative (IDIA), Faculty of Physical and Mathematical Sciences, University of Chile, Chile.}
\affiliation{Millennium Institute of Astrophysics (MAS), Nuncio Monse\~nor S\'otero Sanz 100, Providencia, Santiago, Chile}
\affiliation{Center for Mathematical Modeling (CMM), Universidad de Chile, Beauchef 851, Santiago 8320000, Chile.}

\author[0000-0002-5605-2219]{Anna Franckowiak} \affiliation{Ruhr University Bochum, Faculty of Physics and Astronomy, Astronomical Institute (AIRUB), Universit\"{a}tsstra{\ss}e 150, 44801 Bochum, Germany}

\author[0000-0001-9553-4723]{Christopher Frohmaier} \affiliation{School of Physics and Astronomy, University of Southampton, Southampton, SO17 1BJ, UK}

\author[0000-0002-1296-6887]{Llu\'{i}s Galbany} \affiliation{Institute of Space Sciences (ICE-CSIC), Campus UAB, Carrer de Can Magrans, s/n, E-08193 Barcelona, Spain.} \affiliation{Institut d'Estudis Espacials de Catalunya (IEEC), 08860 Castelldefels (Barcelona), Spain}

\author[0000-0001-6065-5218]{Renato~B.~Galleguillos} \affiliation{Faculty of Engineering, Finis Terrae University, 1509 Pedro de Valdivia Ave, Providencia, RM, Chile} \affiliation{Millennium Institute of Astrophysics (MAS), Nuncio Monse\~nor S\'otero Sanz 100, Providencia, Santiago, Chile}

\author[0000-0002-4163-4996]{Ariel Goobar} \affiliation{Oskar Klein Centre, Department of Physics, Stockholm University, SE-10691 Stockholm, Sweden}

\author[0000-0003-2375-2064]{Claudia~P.~Guti\'errez } \affiliation{Institut d'Estudis Espacials de Catalunya (IEEC), 08860 Castelldefels (Barcelona), Spain} \affiliation{Institute of Space Sciences (ICE-CSIC), Campus UAB, Carrer de Can Magrans, s/n, E-08193 Barcelona, Spain.} 

\author[0000-0002-3841-380X]{Saarah~Hall} \affiliation{Department of Physics and Astronomy, Northwestern University, 2145 Sheridan Rd, Evanston, IL 60208, USA} \affiliation{Center for Interdisciplinary Exploration and Research in Astrophysics (CIERA), Northwestern University, 1800 Sherman Ave, Evanston, IL 60201, USA}

\author[0000-0002-5698-8703]{Erica Hammerstein} \affiliation{Department of Astronomy, University of California, Berkeley, CA 94720-3411, USA}

\author[0000-0001-6718-2978]{Kenneth~R.~Herner}\affiliation{Fermi National Accelerator Laboratory, Batavia, IL 60510, USA}

\author[0000-0002-2960-978X]{Isobel ~M.~Hook} \affiliation{Department of Physics, Lancaster University, Lancaster LA1 4YB, UK}

\author[0000-0003-4591-3201]{Macy~J.~Huston}\affiliation{Department of Astronomy, University of California, Berkeley, CA 94720-3411, USA}

\author[0000-0001-5975-290X]{Joel~Johansson} \affiliation{Oskar Klein Centre, Department of Physics, Stockholm University, SE-10691 Stockholm, Sweden}

\author[0000-0002-5740-7747]{Charles~D.~Kilpatrick}\affiliation{Center for Interdisciplinary Exploration and Research in Astrophysics (CIERA), Northwestern University, 1800 Sherman Ave, Evanston, IL 60201, USA}

\author[0000-0001-6315-8743]{Alex~G.~Kim} \affiliation{Lawrence Berkeley National Laboratory, 1 Cyclotron Road, MS 50B-4206, Berkeley, CA 94720, USA}

\author[0000-0002-3803-1641]{Robert~A.~Knop} \affiliation{Lawrence Berkeley National Laboratory, 1 Cyclotron Road, MS 50B-4206, Berkeley, CA 94720, USA}

\author[0000-0001-8594-8666]{Marek~P.~ Kowalski}\affiliation{Deutsches Elektronen Synchrotron DESY, Platanenallee 6, 15738 Zeuthen, Germany} \affiliation{Institut f\"{u}r Physik, Humboldt-Universit\"{a}t zu Berlin, 12489 Berlin, Germany}

\author[0000-0003-3108-1328]{Lindsey~A.~Kwok} \affiliation{Center for Interdisciplinary Exploration and Research in Astrophysics (CIERA), Northwestern University, 1800 Sherman Ave, Evanston, IL 60201, USA}

\author[0000-0002-2249-0595]{Natalie~LeBaron} \affiliation{Department of Astronomy, University of California, Berkeley, CA 94720-3411, USA}

\author[0000-0001-8967-2281]{Kenneth W. Lin} \affiliation{Department of Astronomy, University of California, Berkeley, CA 94720-3411, USA} \affiliation{Lawrence Berkeley National Laboratory, 1 Cyclotron Road, MS 50B-4206, Berkeley, CA 94720, USA}

\author[0000-0002-7866-4531]{Chang~Liu} \affiliation{Department of Physics and Astronomy, Northwestern University, 2145 Sheridan Rd, Evanston, IL 60208, USA} \affiliation{Center for Interdisciplinary Exploration and Research in Astrophysics (CIERA), Northwestern University, 1800 Sherman Ave, Evanston, IL 60201, USA}

\author[0000-0001-9611-0009]{Jessica~R.~Lu} 
\affiliation{Department of Astronomy, University of California, Berkeley, CA 94720-3411, USA}

\author[0000-0002-1568-7461]{Wenbin Lu} 
\affiliation{Department of Astronomy, University of California, Berkeley, CA 94720-3411, USA}
\affiliation{Theoretical Astrophysics Center, UC Berkeley, Berkeley, CA 94720, USA}

\author[0000-0001-9454-4639]{Ragnhild~Lunnan}\affiliation{Oskar Klein Centre, Department of Astronomy, Stockholm University, Albanova University Center, 106 91 Stockholm, Sweden}

\author[0000-0002-9770-3508]{Kate Maguire} \affiliation{School of Physics, Trinity College Dublin, College Green, Dublin 2, Ireland} 

\author[0000-0002-7466-4868]{Lydia Makrygianni} \affiliation{Department of Physics, Lancaster University, Lancaster LA1 4YB, UK}

\author[0000-0003-4768-7586]{Raffaella~Margutti} \affiliation{Department of Astronomy, University of California, Berkeley, CA 94720-3411, USA} \affiliation{Department of Physics, University of California, 366 Physics North MC 7300, Berkeley, CA 94720, USA}

\author{Dan~Maoz} \affiliation{School of Physics and Astronomy, Tel Aviv University, Tel Aviv 69978, Israel}

\author[0000-0002-9553-2987]{Patrik~Mil\'an~Veres} \affiliation{Ruhr University Bochum, Faculty of Physics and Astronomy, Astronomical Institute (AIRUB), Universit\"{a}tsstra{\ss}e 150, 44801 Bochum, Germany}

\author[0000-0001-8385-3727]{Thomas Moore}   \affiliation{Astrophysics Research Centre, School of Mathematics and Physics, Queens University Belfast, Belfast BT7 1NN, UK}

\author[0000-0002-8070-5400]{A.~J.~Nayana} \affiliation{Department of Astronomy, University of California, Berkeley, CA 94720-3411, USA}

\author[0000-0002-2555-3192]{Matt~Nicholl} \affiliation{Astrophysics Research Centre, School of Mathematics and Physics, Queens University Belfast, Belfast BT7 1NN, UK}

\author[0000-0001-8342-6274]{Jakob Nordin}  \affiliation{Institut f\"{u}r Physik, Humboldt-Universit\"{a}t zu Berlin, 12489 Berlin, Germany}

\author[0000-0003-0006-0188]{Giuliano Pignata} \affiliation{Instituto de Alta Investigaci{\'{o}}n, Universidad de Tarapac{\'{a}}, Casilla 7D, Arica, Chile} \affiliation{Millennium Institute of Astrophysics (MAS), Nuncio Monse\~nor S\'otero Sanz 100, Providencia, Santiago, Chile}

\author[0000-0002-1633-6495]{Abigail~Polin} \affiliation{{Purdue University, Department of Physics and Astronomy, 525 Northwestern Ave, West Lafayette, IN 4790720, USA}}

\author[0000-0003-1470-7173]{Dovi~Poznanski} \affiliation{School of Physics and Astronomy, Tel Aviv University, Tel Aviv 69978, Israel} \affiliation{Cahill Center for Astrophysics, California Institute of Technology, Pasadena CA 91125, USA} \affiliation{Kavli Institute for Particle Astrophysics and Cosmology, Stanford University, 452 Lomita Mall, Stanford, CA 94305, USA} \affiliation{Department of Physics, Stanford University, 382 Via Pueblo Mall, Stanford, CA 94305, USA}

\author[0000-0003-1072-2712]{Jose~L.~Prieto} \affiliation{Instituto de Estudios Astrof\'{i}sicos, Facultad de Ingenier\'{i}a y Ciencias, Universidad Diego Portales, Avenida Ejercito Libertador 441, Santiago, Chile.} \affiliation{Millennium Institute of Astrophysics (MAS), Nuncio Monse\~nor S\'otero Sanz 100, Providencia, Santiago, Chile}

\author{David~L.~Rabinowitz}\affiliation{Department of Physics, Yale University, P.O. Box 208120, New Haven, CT 06520-8101, USA}

\author[0000-0002-5683-2389]{Nabeel Rehemtulla} \affiliation{Department of Physics and Astronomy, Northwestern University, 2145 Sheridan Rd, Evanston, IL 60208, USA}
\affiliation{Center for Interdisciplinary Exploration and Research in Astrophysics (CIERA), Northwestern University, 1800 Sherman Ave, Evanston, IL 60201, USA}
\affiliation{NSF-Simons AI Institute for the Sky (SkAI), 172 E. Chestnut St., Chicago, IL 60611, USA}

\author[0000-0002-8121-2560]{Mickael~Rigault} \affiliation{Universite Claude Bernard Lyon 1, CNRS, IP2I Lyon/IN2P3, UMR 5822, F-69622, Villeurbanne, France}

\author[0000-0002-4429-3429]{Dan~Ryczanowski} \affiliation{Institute of Cosmology and Gravitation, University of Portsmouth, Burnaby Road, Portsmouth, PO1 3FX, UK} \affiliation{School of Physics and Astronomy, University of Birmingham, Birmingham, B15 2TT, United Kingdom}

\author[0000-0003-2700-1030]{Nikhil Sarin} \affiliation{Oskar Klein Centre, Department of Physics, Stockholm University, SE-10691 Stockholm, Sweden} \affiliation{Nordita, Stockholm University and KTH Royal Institute of Technology, Hannes Alfv\'{e}ns v\:{a}g 12, SE-106 91 Stockholm, Sweden}

\author[0000-0001-6797-1889]{Steve~Schulze} \affiliation{Center for Interdisciplinary Exploration and Research in Astrophysics (CIERA), Northwestern University, 1800 Sherman Ave, Evanston, IL 60201, USA}

\author[0009-0009-1590-2318]{Ved~G.~Shah} \affiliation{Department of Physics and Astronomy, Northwestern University, 2145 Sheridan Rd, Evanston, IL 60208, USA}
\affiliation{Center for Interdisciplinary Exploration and Research in Astrophysics (CIERA), Northwestern University, 1800 Sherman Ave, Evanston, IL 60201, USA}
\affiliation{NSF-Simons AI Institute for the Sky (SkAI), 172 E. Chestnut St., Chicago, IL 60611, USA}

\author[0000-0002-6527-1368]{Xinyue Sheng}\affiliation{Astrophysics Research Centre, School of Mathematics and Physics, Queens University Belfast, Belfast BT7 1NN, UK}

\author[0009-0009-4622-7749]{Samuel~P.~R.~Shilling}\affiliation{Department of Physics, Lancaster University, Lancaster LA1 4YB, UK}

\author[0000-0001-5882-3323]{Brooke~D.~Simmons} \affiliation{Department of Physics, Lancaster University, Lancaster LA1 4YB, UK}

\author[0000-0003-2091-622X]{Avinash Singh} \affiliation{Oskar Klein Centre, Department of Astronomy, Stockholm University, Albanova University Center, 106 91 Stockholm, Sweden} \affiliation{Hiroshima Astrophysical Science Centre, Hiroshima University, 1-3-1 Kagamiyama, Higashi-Hiroshima, Hiroshima 739-8526, Japan}

\author[0000-0003-4494-8277]{Graham~P.~Smith}\affiliation{School of Physics and Astronomy, University of Birmingham, Birmingham, B15 2TT, United Kingdom}\affiliation{Department of Astrophysics, University of Vienna, T\"urkenschanzstrasse 17, 1180 Vienna, Austria}

\author[0000-0002-3321-1432]{Mathew~Smith}\affiliation{Department of Physics, Lancaster University, Lancaster LA1 4YB, UK}

\author[0000-0003-1546-6615]{Jesper~Sollerman} \affiliation{Oskar Klein Centre, Department of Astronomy, Stockholm University, Albanova University Center, 106 91 Stockholm, Sweden}

\author[0000-0001-6753-1488]{Maayane~T.~Soumagnac} \affiliation{Department of Physics, Bar-Ilan University, Ramat-Gan 52900, Israel} \affiliation{Lawrence Berkeley National Laboratory, 1 Cyclotron Road, MS 50B-4206, Berkeley, CA 94720, USA}

\author[0000-0003-0347-1724]{Christopher~W.~Stubbs} \affiliation{Department of Physics, Harvard University, 17 Oxford St, Cambridge, MA 02138, USA} \affiliation{Department of Astronomy, Harvard University, 60 Garden St, Cambridge MA, 02138, USA}

\author[0000-0001-9053-4820]{Mark~Sullivan} \affiliation{School of Physics and Astronomy, University of Southampton, Southampton, SO17 1BJ, UK}

\author[0009-0005-8230-030X]{Aswin~Suresh} \affiliation{Department of Physics and Astronomy, Northwestern University, 2145 Sheridan Rd, Evanston, IL 60208, USA} \affiliation{Center for Interdisciplinary Exploration and Research in Astrophysics (CIERA), Northwestern University, 1800 Sherman Ave, Evanston, IL 60201, USA}

\author[0000-0002-3683-7297]{Benny Trakhtenbrot} \affiliation{School of Physics and Astronomy, Tel Aviv University, Tel Aviv 69978, Israel}

\author[0000-0002-4557-6682]{Charlotte Ward} \affiliation{Department of Astrophysical Sciences, Princeton University, Princeton, NJ 08544, USA}

\author[0009-0002-4843-2913]{Eli~Wiston} \affiliation{Department of Astronomy, University of California, Berkeley, CA 94720-3411, USA}

\author{Helen~Xiong}\affiliation{Department of Astronomy, Yale University, P.O. Box 208101, New Haven, CT 06520-8101, USA}

\author[0000-0001-6747-8509]{Yuhan Yao}\affiliation{Miller Institute for Basic Research in Science, 468 Donner Lab, Berkeley, CA 94720, USA} \affiliation{Department of Astronomy, University of California, Berkeley, CA 94720-3411, USA}

\author[0000-0002-3389-0586]{Peter~E.~Nugent} \affiliation{Lawrence Berkeley National Laboratory, 1 Cyclotron Road, MS 50B-4206, Berkeley, CA 94720, USA} \affiliation{Department of Astronomy, University of California, Berkeley, CA 94720-3411, USA}



\begin{abstract}

We present the La Silla Schmidt Southern Survey (LS4), a new wide-field, time-domain survey to be conducted with the 1\,m ESO Schmidt telescope. The 268 megapixel LS4 camera mosaics 32 2k$\times$4k fully depleted CCDs, providing a $\sim$20\,deg$^2$ field of view with $1''$ pixel$^{-1}$ resolution. The LS4 camera will have excellent performance at longer wavelengths: in a standard 45\,s exposure the expected 5$\sigma$ limiting magnitudes in $g$, $i$, $z$ are $\sim$21.5, $\sim$20.9, and $\sim$20.3\,mag (AB), respectively. The telescope design requires a novel filter holder that fixes different bandpasses over each quadrant of the detector. Two quadrants will have $i$ band, while the other two will be $g$ and $z$ band and color information will be obtained by dithering targets across the different quadrants. The majority (90\%) of the observing time will be used to conduct a public survey that monitors the extragalactic sky at both moderate (3\,d) and high (1\,d) cadence, as well as focused observations within the Galactic bulge and plane. Alerts from the public survey will be broadcast to the community via established alert brokers. LS4 will run concurrently with the Vera C.~Rubin Observatory's Legacy Survey of Space and Time (LSST). The combination of LS4$+$LSST will enable detailed holistic monitoring of many nearby transients: high-cadence LS4 observations will resolve the initial rise and peak of the light curve while less-frequent but deeper observations by LSST will characterize the years before and after explosion.  Here, we summarize the primary science objectives of LS4 including microlensing events in the Galaxy, extragalactic transients, the search for electromagnetic counterparts to multi-messenger events, and cosmology. 

\end{abstract}



\section{Introduction} \label{sec:intro}

The quest to map the sky at high precision and to great depths has resulted in an unprecedented proliferation of wide-field, optical time-domain surveys. The need for repeated observations to build depth results in the production of light curves even when the discovery of transients and variables is not the primary science objective of a given experiment. This push toward the time domain has been driven by a bevy of exciting new discoveries in the past $\sim$two decades, including significant advances in multi-messenger astronomy with the discovery of an optical counterpart to a gravitational wave (GW) event \citep{Abbott17a} and new neutrino sources being associated with electromagnetic radiation \citep[e.g.,][]{IceCube-Collaboration18,Stein21}. The spread of time-domain surveys ranges from bespoke efforts on small aperture telescopes to study very specific phenomena to extremely large projects designed to study the cosmos both with the time variable information they capture and the very deep images they produce by combining individual exposures. This later strategy is best illustrated by the forthcoming Legacy Survey of Space and Time \citep[LSST;][]{Ivezic19} conducted by the Vera C.~Rubin Observatory, which will provide a decade-long movie capturing the variability of faint sources in the southern hemisphere on timescales ranging from days to years. 

Many past and on-going experiments have laid the foundation for LSST by characterizing the time-variable sky, albeit at shallower depths or more narrow areas. These projects include: the Optical Gravitational Lensing Experiment \citep[OGLE;][]{Udalski92}, the All-sky Automated Survey \citep[ASAS;][]{Pojmanski97}, Palomar-QUEST \citep{Djorgovski08}, the Catalina Sky Survey \citep[CSS;][]{Larson03} and its associated Catalina Real-time Transient Survey \citep[CRTS;][]{drake2009}, Sloan Digital Sky Survey-II \citep[SDSS-II;][]{Frieman08}, Skymapper \citep{Keller07}, PanSTARRS \citep{Kaiser10}, the Palomar Transient Factory \citep[PTF;][]{law09}, the All-sky Automated Survey for Supernovae \citep[ASAS-SN;][] {Shappee14}, the Evryscope \citep{Law15}, the Dark Energy Survey \citep[DES;][]{Dark-Energy-Survey-Collaboration16}, the Asteroid Terrestrial-impact Last Alert System \citep[ATLAS;][]{Tonry18}, the Zwicky Transient Facility \citep[ZTF;][]{Bellm19}, the Young Supernova Experiment \citep[YSE;][]{Jones21}, BlackGEM \citep{Bloemen16}, and the Gravitational-Wave Optical Transient Observer \citep[GOTO;][]{Steeghs22}. Collectively these projects have developed new insights for sources as close as moving objects within our solar system to distant supermassive black holes (SMBHs) that are more than halfway across the visible Universe. Collectively, these efforts have motivated and influenced the design of future upcoming surveys. 

In this paper, we present a new time-domain survey, the La Silla Schmidt Southern Survey (LS4). LS4 is designed to discover extragalactic explosions and variable sources within the Milky Way using the La Silla Schmidt telescope along with the LS4 camera (see Section~\ref{sec:tel+camera}). Unlike the time-domain surveys listed above, LS4 has been designed to run concurrently with LSST, providing incredible opportunities for synergy (Section~\ref{sec:lsst-synergy}). LS4 can survey wider areas at a higher cadence than LSST while also monitoring the bright transients that saturate the LSST detector, LSSTCam. These bright transients are especially valuable as they are the most amenable to spectroscopic follow-up and multi-wavelength investigations, and therefore have an outsized role in advancing our understanding of extragalactic explosions. Below, we summarize the technical details of the telescope and camera, outline the LS4 survey strategy, and discuss the science that LS4 can achieve on its own while also describing the results that will be possible by combining LSST observations with the LS4 survey.

\section{Overview of the ESO Schmidt, LS4 Camera, and LS4 Transient Discovery Pipeline}\label{sec:tel+camera}

LS4 will be conducted using the ESO 1\,m Schmidt Telescope located at the La Silla Observatory in Chile. The LS4 partnership will lease the Schmidt from ESO for a five year period to conduct the survey. The development of the LS4 camera and survey operations are paid for via funds raised by the partnership. The ESO Schmidt has a long history of conducting wide-field surveys, most recently as part of the La Silla-QUEST Survey \citep{Rabinowitz12,Baltay13}, and LS4 will build on that legacy.

\begin{figure*}[tbph]
    \centering
    \includegraphics[width=\textwidth]{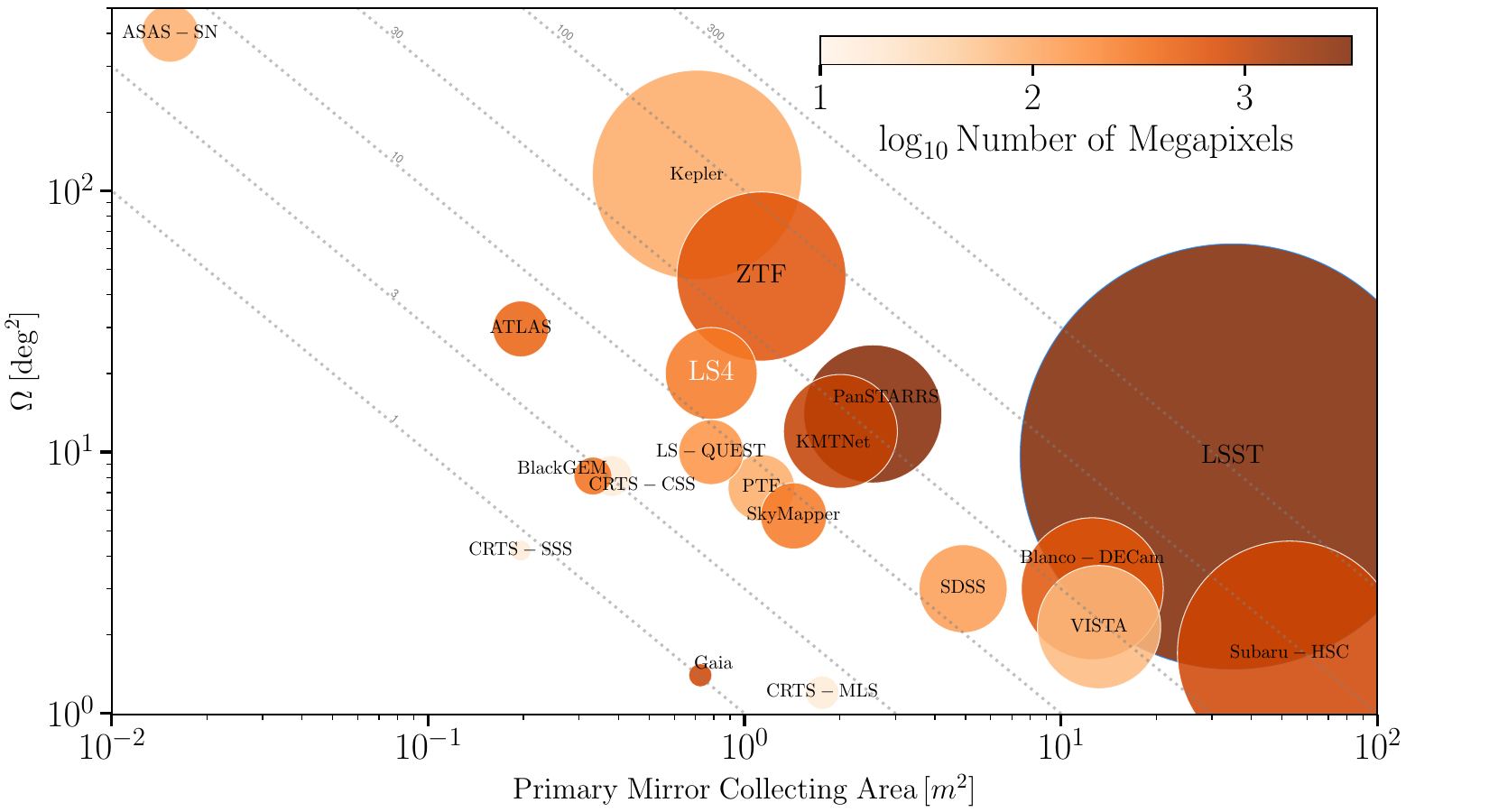}
    \caption{A comparison of the FOV ($\Omega$) and light collecting area of the primary mirror ($A$) highlighting the \'etendue ($A\Omega$) of past, current, and future time-domain surveys. The radius of the symbols is proportional to the \'etendue while the color corresponds to the total number of megapixels for the system. For surveys that employ multiple telescopes (ASAS-SN, ATLAS, Gaia, KMTNet, and PanSTARRS), the primary mirror collecting area is reported for a single telescope, while the FOV corresponds to the FOV of a single unit multiplied by the total number of telescopes (i.e., the figure reflects the optimistic scenario that every telescope is operational simultaneously). Dashed lines of constant \'etendue are shown along the diagonal. Among 1\,m class telescopes, LS4 has a larger \'etendue than all but Kepler and ZTF.
    }\label{fig:etendue}
\end{figure*}

\subsection{The LS4 Camera}

\begin{figure}[tbph]
    \centering
    \includegraphics[width=3.2in]{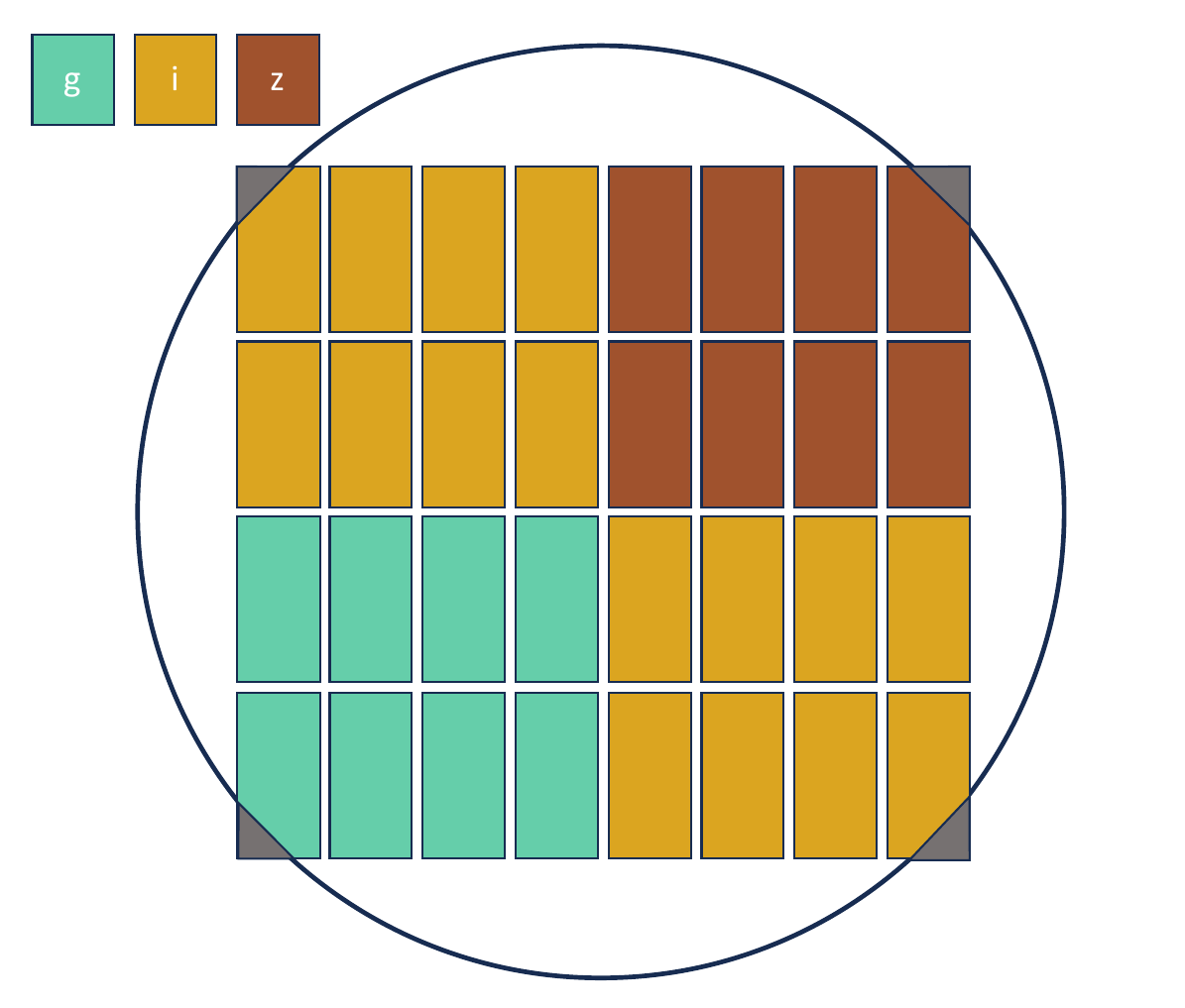}
    \caption{A schematic of the LS4 filter design on the CCD focal plane. The 32 CCDs are represented by the rectangles arranged in an array and shaded according to the filter that they are beneath at the start of the survey. The detectors located at the corners are partially occulted by the edges of the field-flattener lens, which is outlined as the circle.
    }\label{fig:filterlayout}
\end{figure}

Full details of the LS4 camera and its performance are reported in Lin et al.\ (2025, in prep.). Briefly, LS4 leverages an upgrade to the QUEST camera \citep{Baltay07}, which was used to conduct the La Silla-QUEST Survey, by replacing the detectors on the focal plane with 32 science-grade, 2k $\times$ 4k fully depleted CCDs that were fabricated for, but ultimately not installed on, the Dark Energy Camera \citep[DECam;][]{Flaugher12}. The 268-megapixel camera completely fills the Schmidt field of view (FOV), delivering $\sim$20\,$\mathrm{deg}^{2}$ images. With the notable exception of the Kepler satellite \citep{Borucki10} and ZTF, LS4 has a larger \'etendue than any other time-domain survey with a similar aperture size as shown in Figure~\ref{fig:etendue}. Each DECam CCD on the camera has two amplifiers located at two corners, and can be read out in two modes: dual amplifier (with a read-out time of $\sim$17\,s) and single amplifier (read-out time of $\sim$34\,s). When read out is in dual amplifier mode, four amplifiers fail to produce signal, reducing the overall FOV by $\sim$6\% (4/64). There is no room for a filter wheel within the telescope, and thus LS4 employs a novel design wherein four different filters, each covering a different quadrant of the camera, are affixed within the optical path of the detector. This design is illustrated in Figure~\ref{fig:filterlayout}, and at the start of the survey, there are two $i$-band filters, one $g$-band filter, and one $z$-band filter occupying the four quadrants.\footnote{The LS4 collaboration also has an $r$-band filter, which could in the future be swapped in for any of the current filters in front of the focal plane.} During survey operations, color information will be obtained by dithering the telescope across different quadrants while pointing at the same locations in the sky. The LS4 filters were designed by Asahi Spectra using the same design as the filters being used with LSSTCam for LSST (the filter response curves are reported in Lin et al.\ 2025, in prep.). At the joints between two filters, approximately 2\% of the total number of available imaging pixels are occulted by the filter holders, with incident photons entering from multiple filters.  

\subsection{The Difference Imaging Detection Pipeline}

Transients from LS4 will be detected in near real-time using the difference imaging pipeline \texttt{SeeChange}\footnote{\url{https://github.com/c3-time-domain/SeeChange}} (see Lin et al.~2025, in prep. for additional details). The LS4 pipeline utilizes a ``conductor'', a component running on the National Energy Research Scientific Computing Center (NERSC) Spin kubernetes cluster \citep{Snavely2018} that monitors files from the telescope and tracks which exposures need to be processed.  Processes running on the NERSC compute nodes will regularly contact the conductor to be assigned exposures to process. The pipeline splits an exposure into its individual CCDs and launches parallel processes (one for each CCD) to simultaneously process all the CCDs.  Standard procedures are used to identify variables and transients: pre-processing (overscan correction, flat fielding, etc.), source extraction, point spread function (PSF) estimation, astrometric calibration, photometric calibration, the subtraction of a reference image, and finally, residual identification.  
Only detections above a signal-to-noise ratio threshold that pass simple preliminary cuts, designed to eliminate dipoles (not too many negative pixels), and eliminate residuals that are substantially different than the PSF, are saved.  It then creates 51-pixel thumbnails of the science, template, and difference images around surviving detections, and runs these thumbnails through the \texttt{RBbot} (Rehemtulla et al.~2025, in prep.) to produce a real/bogus score \citep[see e.g.,][]{Bloom12} between 0 and 1.  

The pipeline generates alerts for all candidates that pass the preliminary filters (see Lin et al.\ 2025, in prep.\ for the full details).  One alert encapsulates one detection.  The alert schema\footnote{\url{https://github.com/c3-time-domain/SeeChange/tree/ls4_production/share/avsc}} are based on, but modified from and greatly stripped down from, LSST alert schema.  Alerts include basic image metadata (time of the exposure, filter, measurement of the seeing, measurement of the limiting magnitude), flux measurements, the real/bogus score, 51-pixel thumbnails around the detection from the science, template, and difference images, and summaries of any previous detections by the pipeline of a transient at the same position on the sky.  The pipeline is capable of distributing alerts to multiple kafka servers.  Public alerts from LS4 will be distributed through SCiMMA\footnote{\url{https://scimma.org}}, which will sit upstream from the brokers and individuals that will ingest LS4 alerts.  

For initial object subtraction, astrometric solution, PSF estimation, image alignment, and reference coaddition, \texttt{SeeChange} by default uses the Astromatic tools \texttt{SExtractor}, \texttt{SCAMP}, and \texttt{Swarp} \citep{bertin96}.  Image subtraction currently uses either the ZOGY \citep{Zackay16} or Alard/Lupton \citep{Alard98} algorithms.  Astrometric solutions and photometric calibration are both performed relative to Gaia DR3 stars \citep{Gaia-Collaboration23}.  

A unique aspect of \texttt{SeeChange} is that it is designed to run the same data through the pipeline multiple times using different code versions or different sets of parameters.  It stores the results of every run in the database; updating code or parameters does not require removing previous results from the database, or re-initializing the database.  \texttt{SeeChange} tracks a ``provenance'' of every data product, which includes the name of the process (e.g., ``astrometric calibration'' or ``subtraction''), the code version of that process, the parameters used to run the process, and links to the provenances of any upstream processes.  This provenance system allows tracking of processing and detection through the inevitable code changes and parameter tuning that will occur especially during the early part of the survey.  Users will be able to understand whether and why a specific transient was or was not detected while the survey was running, and analyses of detection efficiency will be able to ensure consistent processing where that is necessary.

\section{The LS4 Survey}\label{sec:ls4_survey}

With its \fov\ FOV, LS4 will conduct a rapid time-domain survey of the southern sky. As discussed in more detail below (see Section~\ref{sec:ls4_science}), the LS4 survey is designed to discover transients and variables while increasing the overall scientific output of the project by leveraging ongoing observations from other surveys that will also have coverage in the southern hemisphere, such as LSST \citep{Ivezic19}, ZTF \citep{Bellm19,Graham19}, BlackGEM \citep{Bloemen16}, Euclid \citep{Laureijs11}, ULTRASAT \citep{Shvartzvald24}, the Ultraviolet Explorer \citep[UVEX;][]{Kulkarni21}, and the Nancy Grace Roman Space Telescope \citep{Akeson19}. The vast majority of the LS4 observing time (90\%) will be used to conduct a public survey, while the remaining 10\% will be used by the LS4 collaboration for special projects. 

\subsection{The LS4 public survey}

The LS4 public survey is primarily concentrated away from the Galactic plane to maximize the discovery of extragalactic transients and variables. Fields with Galactic latitude $|b| \geq 10^\circ$ are considered extragalactic in LS4. A subset of observations will target the Milky Way to search for stellar variables and transient events, such as microlensing. Thus, the LS4 public survey is divided into three major campaigns covering (i) a wide area of the extragalactic sky, (ii) a small area to be observed at a higher cadence, and (iii) the Galactic plane. All observations from the public survey will be automatically processed and distributed to alert brokers for use by the community. 

\subsubsection{The LS4 Long-cadence ExtraGalactic (LEG) Survey }\label{sec:survey:leg}

Roughly 50\% of the public survey will be devoted to the LS4 Long-cadence ExtraGalactic (LEG) Survey. The LEG Survey will monitor, on average, a $\sim$7,000\,$\deg^{2}$ area with a 3-day cadence in the $g$, $i$, and $z$ filters. Fields within the LEG footprint will be observed twice per night with a minimum $\sim$30\,min separation between images to identify and reject Solar system moving objects from the extragalactic transient alert stream. Following the success of the ZTF Northern Sky Survey \citep{Bellm19a}, the dual nightly visits will be conducted in different filters by shifting the telescope pointing by half the FOV between visits. On a given night, a source will be observed in either the $g$+$i$ filters or the $i$+$z$ filters, and when the field is revisited 3 days later, it will be observed with the opposite filter complement (i.e., the 3-day cadence refers to revisit time in the $i$-band). LS4 LEG will actively monitor both ``high season'' and ``low season'' LSST fields, once rolling observations of the Wide, Fast, Deep \citep[WFD;][]{Bianco+2022} survey begin. LS4 LEG will discover and characterize long-lived transients over the entirety of the southern extragalactic sky. 

\subsubsection{The LS4 Fast Observations of Optical Transients (FOOT) Survey}\label{sec:survey:foot}

Roughly 25\% of the public survey will be devoted to the LS4 Fast Observations of Optical Transients (FOOT) Survey. LS4 FOOT will monitor, on average, a 1200\,$\deg^{2}$ area with a 1-day cadence using the same strategy as LS4 LEG, that is, observations will be collected in the $i$-band every night, while $g$- and $z$-band observations will alternate every other night. High-cadence LS4 FOOT observations will specifically focus on rapidly evolving transients and primarily operate in LSST WFD ``low season'' fields. This observational strategy will ensure that Rubin captures very deep imaging in the year before and the year after new LS4 FOOT discoveries are made.

Intrasurvey cadence between LS4, Rubin, and other public optical time-domain surveys (e.g.,  ATLAS and ZTF) covering the same footprint will enable a range of science cases covering intrinsically fast and young transients. By strategically obtaining LS4 observations on timescales of 1~day \textit{after} scheduled Rubin observations, we will maximize the number of young transients whose rising light curves cross the threshold of LS4's $\approx$21~mag limiting survey magnitude and guarantee detections in both surveys. This strategy has yielded several SN detections within 2~days of explosion in YSE using Pan-STARRS in conjunction with ZTF \citep[see, e.g.,][and references therein]{Jones21,Terreran22,Aleo+2023,Jacobson-Galan24}, and the increase in survey volume from LS4 and Rubin can expand on these results.

\subsubsection{The LS4 Stellar Oscillations, Lensing, and Eruptions (SOLE) Survey}

The remaining $\sim$25\% of the public survey will be devoted to the LS4 Stellar Oscillations, Lensing, and Eruptions (SOLE) Survey. LS4 SOLE will target high-density star fields, primarily at low-galactic latitudes, in the plane between $-72^\circ \lesssim l \lesssim 36^\circ$ and $|b| \lesssim 4^\circ$, and in the bulge, from $|l|\lesssim 20^\circ$ and $|b| \lesssim 8^\circ$, to identify stellar transients (e.g., microlensing events) and variables. This $\sim1200\,\deg^{2}$ Galactic plane region will be observed with a 1\,d cadence in the $i$-band, with all fields cycling through $g$ and $z$, similar to the LS4 LEG and LS4 FOOT surveys. 

\subsection{The Partnership Survey}

One tenth of the LS4 observing time is reserved for the LS4 partnership. Alerts generated via observations from the partnership survey will only be released to the public after an expected proprietary period of $\sim$30\,d. The fields to be observed and the time budgeted to them from the LS4 partnership survey are determined via an internal proposal process, and thus, the focus can change over time. At the start of survey operations, partnership observations will be primarily focused on observing fields that are regularly monitored by other surveys as well as follow-up of GW events discovered by LIGO-Virgo-KAGRA (LVK). 

\subsubsection{LS4 Deep Fields}

Using partnership time, LS4 will co-observe fields that are regularly monitored by the space-based Euclid and ULTRASAT satellites. These observations will be conducted at a minimum cadence of 1\,d, and form the basis for the LS4 deep fields survey. The LS4 FOV is especially well-matched to the Euclid Deep Field South ($\sim$23\,$\deg^{2}$) and Euclid Deep Field Fornax \citep[$\sim$210\,$\deg^{2}$; see, e.g.,][]{EuclidprepI,EuclidprepXVII}. LS4 observations will provide high-cadence monitoring of Euclid transients. Similar to Rubin, Euclid saturates for moderately bright sources ($\sim$17.8\,mag), meaning LS4 will provide high-precision measurements for the  brightest transients within the Euclid footprint. During the southern winter, ULTRASAT will spend $>$21\,hr day$^{-1}$ staring at a single field to provide extremely high-cadence near-ultraviolet (near-UV) observations of a 204\,$\deg^{2}$ FOV near the southern ecliptic pole \citep{Shvartzvald24}. LS4 will monitor this area with a 1\,d cadence whenever it is visible. High-cadence observations of the Euclid and ULTRASAT deep fields will form the basis for the LS4 deep fields survey.

\subsubsection{Gravitational Wave ToOs}

LS4 target-of-opportunity (ToO) observations will be used to search the wide-area localizations for GW events discovered by LVK as part of the LS4 partnership time. LS4 ToOs for LVK events will be activated for new discoveries that meet the following criteria: (i) the LS4 search area is less than 1000\,$\deg^{2}$, which can include events with a localization $>1000\,\deg^{2}$ if e.g., some of the localization area is in the north and will be observed by other facilities, (ii) the probability that the merger includes a neutron star $P(\mathrm{NS}) > 0.1$, and (iii) the false alarm rate is less than 1\,yr$^{-1}$. Transient candidates identified via LS4 LVK ToO observations will be announced to the community via GCNs and TNS.

\subsection{Synergistic Compatibility with LSST}\label{sec:lsst-synergy}

\begin{figure*}[htbp!]
    \centering
    \includegraphics[width=6.5in]{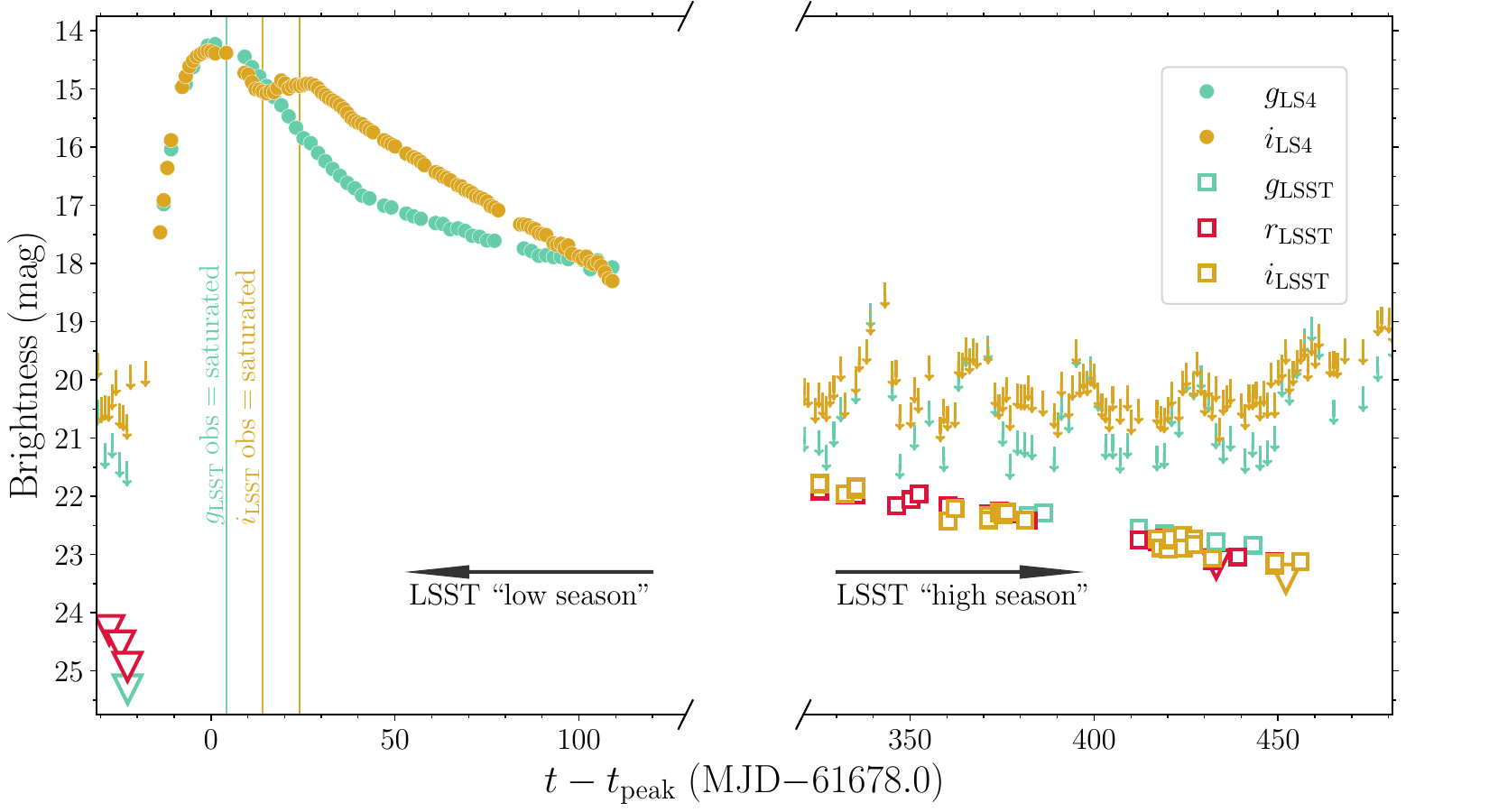}
    \caption{Simulated light curve of a Type Ia supernova based on observations of SN\,2011fe from \citet{Zhang16} shifted to $z = 0.01$, which demonstrates the synergy between LS4 and LSST, especially after LSST begins rolling observations. The LSST WFD survey provides very deep limits a few weeks before $t_{\mathrm{peak}}$ and observations around peak brightness, but the limits are too early to constrain progenitor or explosion models and maximum light observations are saturated. High-cadence LS4 observations place tight constraints on early emission and sample the rise and peak well. During the following observing season, $\sim$1\,yr after explosion, the SN is no longer detected by LS4 but LSST provides dense sampling to monitor the late-time decay as the field enters a WFD ``high season.''. For clarity, uncertainties on the individual observations, the LS4 $z$ band, and the LSST $u$, $z$, and $y$ bands are not shown. LS4 upper limits are shown as downward facing arrows, while LSST upper limits are shown as downward facing triangles.}\label{fig:synergy}
\end{figure*}

The small aperture and large FOV of LS4 are particularly synergistic with the LSST WFD survey. In particular, LSST images will saturate for sources brighter than $r \approx 17.5$\,mag, meaning LS4 can ``take over'' monitoring any bright transients found by LSST that eventually saturate LSSTCam \citep[only exceptionally nearby transients, like SN\,2011fe][will saturate LS4]{Nugent11}. Furthermore, the larger FOV of LS4 allows us to survey at a higher cadence than LSST, especially given the WFD goal to provide nearly uniform depth coverage across the entire southern sky in six different filters. LS4 fields will complement the ``rolling cadence'' to be implemented by WFD, whereby at any given time, roughly half of the visible WFD fields will obtain higher-cadence observations while the other half will be observed at a lower cadence \citep[on average there are $\sim$125 observations yr$^{-1}$ during ``high season'' and only $\sim$25 observations yr$^{-1}$ during ``low season;''][]{PSTN-056}. High-cadence LS4 FOOT observations will be primarily concentrated in WFD low-cadence regions to ensure that LS4 maximally fills the gaps between LSST observations. In Figure~\ref{fig:synergy} we show the simulated light curve of a normal Type Ia supernova (SN\,Ia) at $z = 0.01$, based on observations of SN\,2011fe \citep{Zhang16}, using version 4.3.1 of the baseline simulation of WFD from the Rubin Observatory.\footnote{\url{https://s3df.slac.stanford.edu/data/rubin/sim-data/sims_featureScheduler_runs4.3/baseline/baseline_v4.3.1_10yrs.db}} We make the simplifying assumption that the seeing and cloud cover are perfectly correlated between La Silla, where LS4 is located, and Cerro Pach\'{o}n, where Rubin is located, which is not completely unreasonable given the proximity of the two observatories. The sky noise for the simulated LS4 observations is scaled from the Rubin simulations by the difference in apertures and pixel scales. For the simulated light curves shown below (e.g., in Figure~\ref{fig:synergy}), the simulated transient is located at the arbitrary right ascension and declination $\alpha_\mathrm{J2000} = 00^\mathrm{h} \, 51' \, 38\farcs4, \, \delta_\mathrm{J2000} = -22^\circ \, 32' \, 24\farcs0$. This position is held constant for the different figures to illustrate the differing cadences depending on the LSST observing season. 

As is clear from Figure~\ref{fig:synergy}, focusing LS4 FOOT survey observations in ``low season'' LSST fields provides high-cadence observations when transients are young and bright, the precise epochs when significant evolution occurs on short time scales. At the same time, WFD provides exceptionally deep observations (relative to LS4) in the year before and after the explosion with their own high-cadence observations that can constrain pre-explosion eruptions \citep[e.g.,][]{Jacobson-Galan22} or monitor the late-time decay \citep[e.g.,][]{Graur19} of LS4 FOOT discoveries. Together, LS4 and LSST will systematically monitor the very early and very late evolution of hundreds of new transients.

\subsection{Multi-resolution scene modeling for characterization of LS4 transients} \label{subsec:scene_modeling}

LS4 will additionally leverage higher resolution imaging from LSST and other telescopes to characterize transients with a full scene modeling approach \citep[e.g.,][]{Holtzman95, Brout19}, whereby the transient, the host galaxy, and background galaxies are forward modeled across multi-epoch imaging, enabling a measurement of the transient flux without the need for image differencing. This can be beneficial when combining photometry from multiple surveys without concern for reference image mismatch, measuring relative transient-host nucleus spatial offsets to classify sources as nuclear or non-nuclear, obtaining carefully sampled posteriors for transient and host galaxy parameters, and producing forced photometry of a faint source where the position is not well-constrained \citep{Ward2025}. Given the $1.0^{\prime \prime}$ pixel scale and shallow depth of LS4, the inclusion of high-resolution and deeper images from Rubin or space-based imaging will significantly improve the forward model. 

Joint analysis of LS4 and overlapping ZTF, Rubin, Euclid, Roman, or HST imaging can be done using new scene modeling codes such as \texttt{Scarlet2}: a new version of the \texttt{Scarlet} LSST deblender \citep{Melchior2018} that can use both color and variability information to model variable point sources against a static background across multi-epoch, multi-band, and multi-resolution imaging data. \texttt{Scarlet2} provides the advantages of data-driven priors on galaxy morphologies enabling non-parametric galaxy models and is fully GPU compatible \citep{Sampson2024, Ward2025}.  

\begin{figure}[h]
\begin{subfigure}{}
  \centering
  \includegraphics[width=.95\linewidth]{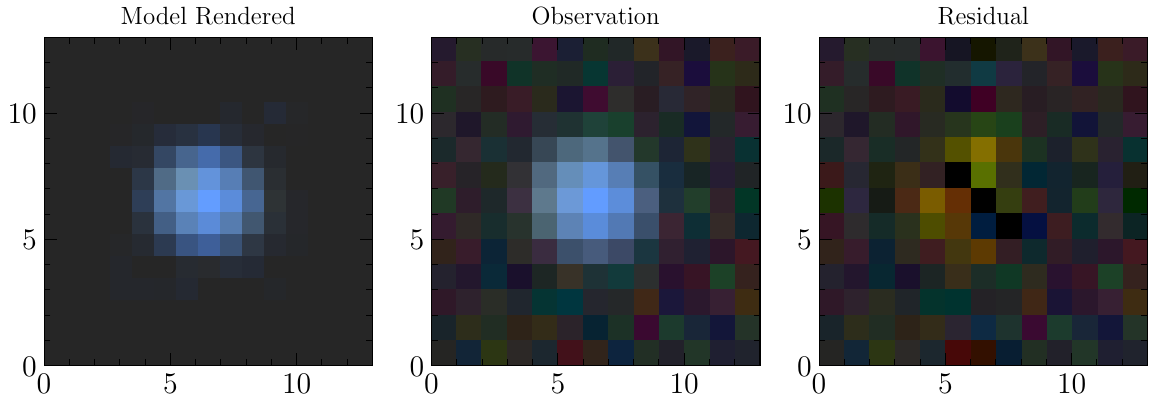}  
\end{subfigure}
\begin{subfigure}{}
  \centering
  \includegraphics[width=.95\linewidth]{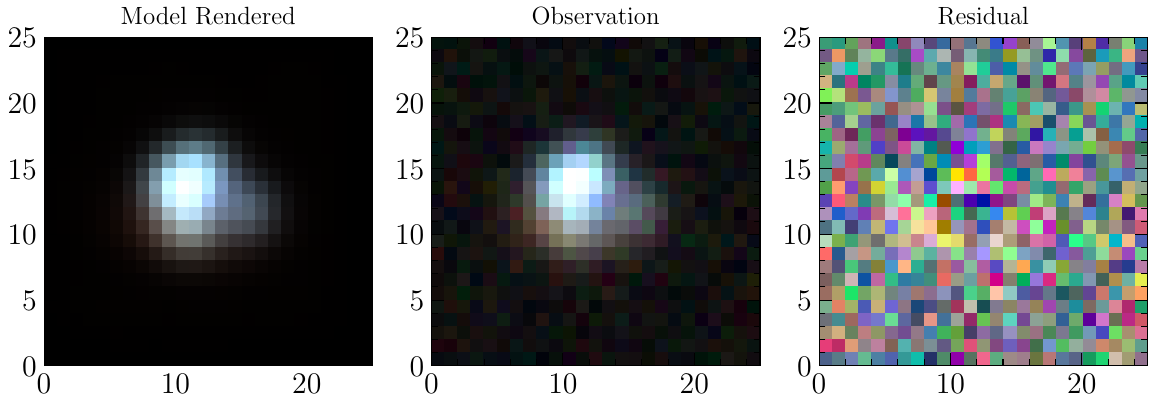}  
\end{subfigure}
\begin{subfigure}{}
  \centering
  \includegraphics[width=.95\linewidth]{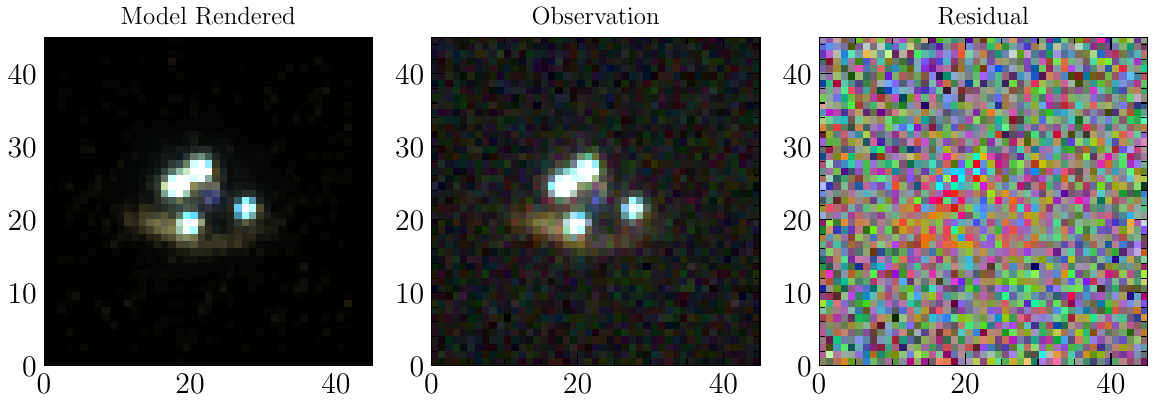}  
  \caption{Example \texttt{Scarlet2} rendered models, observations, and residuals for a gravitationally lensed, quadruply imaged SN jointly modeled across simulated LS4 (\textit{top}), Rubin (\textit{middle}), and Euclid (\textit{bottom}) imaging. We show a $13^{\prime\prime}$ by  $13^{\prime\prime}$ cutout of the LS4 image, and a $5^{\prime\prime}$ by $5^{\prime\prime}$ cutout of the corresponding Rubin and Euclid images. The system is unresolved in LS4 imaging, partially resolved in Rubin imaging, and well-resolved in space-based Euclid imaging. Joint analysis with a higher resolution image enables extraction of an LS4 light curve of each SN image in the system.}\label{fig:multiresgLSNe}
\end{subfigure}

\end{figure}

In Figure~\ref{fig:multiresgLSNe} we show a key LS4 science application where multi-resolution forward modeling will be beneficial: modeling strongly lensed supernovae across LS4, Rubin, and space-based imaging. Pixel-level simulations of the gravitationally lensed supernovae (gLSNe) population detectable by surveys like LS4 and Rubin show that the angular separations of most multiply imaged SNe will range from $\sim0.03-3.0^{\prime \prime}$ with a median of $\sim0.85^{\prime \prime}$, such that most gLSNe will be unresolved or marginally resolved \citep{Goldstein2019}. When LS4 provides the high cadence photometry required to extract precise time delays from multiply-imaged SNe for cosmography, a joint modeling framework will enable us to extract light curves from the poorly resolved systems without the need for a reference image. In another key science application, we will apply multi-resolution scene modeling to measure transient-host spatial offsets, assisting in distinguishing between nuclear and non-nuclear transients, and enabling the identification of spatially offset Tidal Disruption Events (TDEs) from wandering Massive Black Holes (MBHs), as demonstrated in \citet[][]{Yao2025}. Finally, we will use \texttt{Scarlet2} to produce multi-survey light curves for transients that are present in reference imaging, including Active Galactic Nuclei (AGN).

\section{LS4 Science Working Groups}\label{sec:ls4_science}

The analysis of LS4 observations within the collaboration will be conducted by five science working groups (SWGs): (i) Massive Black Holes, which focuses on TDEs, AGN, and other nuclear flares from the center of galaxies, (ii) Physics of Stellar Explosions, focused on core-collapse and thermonuclear SNe, (iii) Multi-messenger Astrophysics, focused on the search for electromagnetic counterparts to GW and neutrino detections, (iv) Cosmology, which will use SNe to study the expansion of the Universe, and (v) Galactic Transients and Variables, focused on stars within the Milky Way. In the sections below we highlight a non-exhaustive list of the science that will be pursued by the individual LS4 SWGs.

\section{Massive Black Holes}

\subsection{Using TDEs to Probe MBH Accretion and Jet Physics} 

When a star passes too close to an MBH, it can be shredded by tidal forces and subsequently accreted \citep{rees88, gezari21}, producing a flare visible from the X-rays to the radio band. These TDEs produce a similar luminosity to SNe in the optical and UV, but they are intrinsically rare. The TDE rate in Milky-Way like galaxies is $\sim 3\times 10^{-5}\,\rm yr^{-1}$ \citep{Yao2023}. Persistent blue optical colors and a long-lasting plateau distinguish TDEs from SNe. A small fraction of TDEs launch relativistic on-axis jets, which can produce a fast transient with an early-time red peak in the optical \citep{Andreoni2022}. 

TDEs provide a laboratory to study the real-time formation and evolution of MBH accretion flows \citep{Wevers2021, Yao2022_21ehb, Yao2024_22lri}, as well as particle acceleration and energy dissipation in newly launched jets \citep{Burrows2011, DeColle2020, Yao2024_22cmc}. Multi-wavelength observations are crucial for future progress, but follow-up campaigns can only begin once a TDE has been discovered. There are few optical TDEs with a well-characterized rise and peak, but high-cadence surveys, such as LS4, will discover more events during this critical phase.  

\begin{figure*}[htbp]
\centering
    \includegraphics[width=6in]{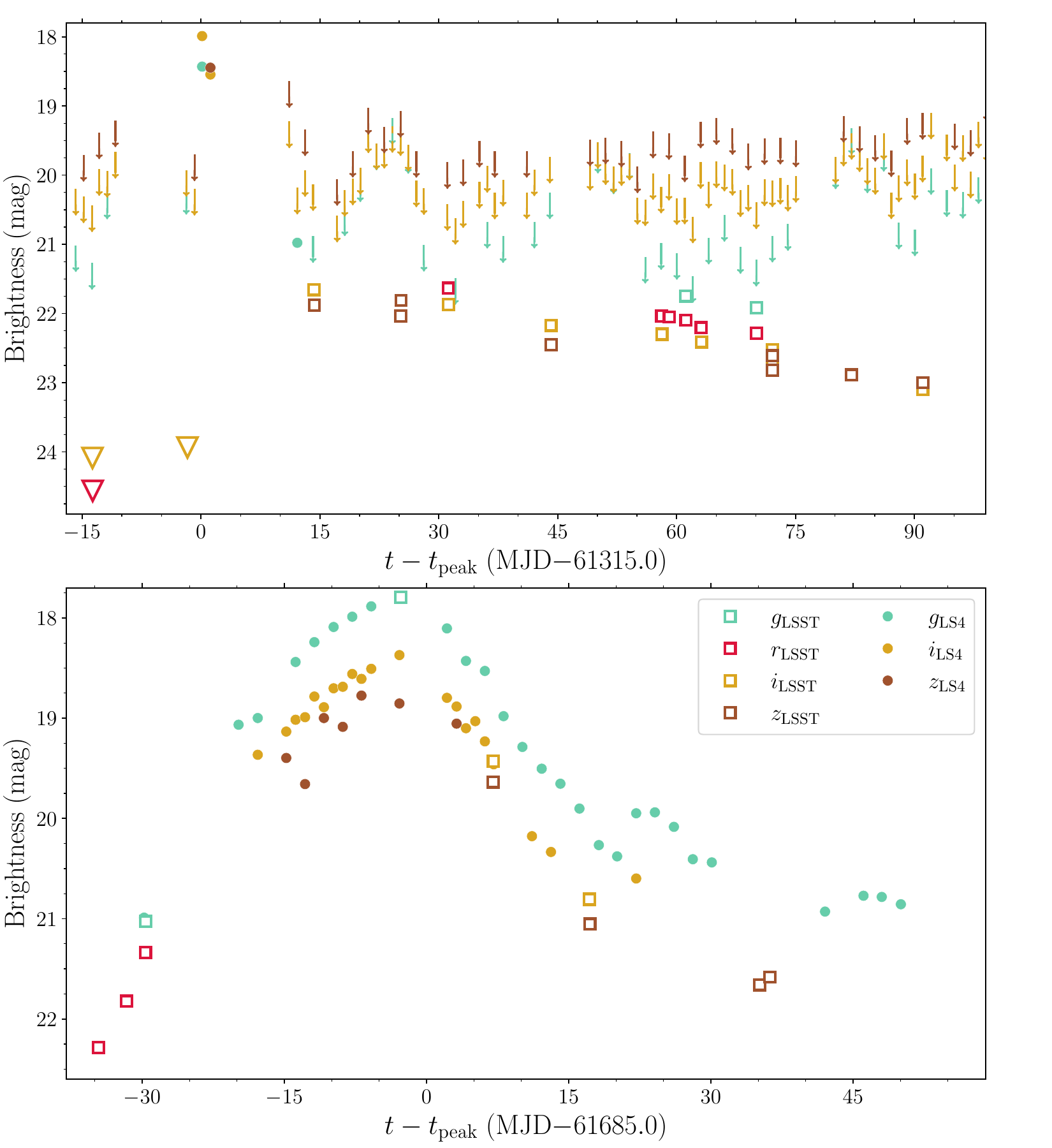}
    \caption{Simulated TDE light curves as observed by LS4 and LSST. \textit{Top}: a jetted TDE at $z=0.9$, using the SED model of AT\,2022cmc (Hammerstein et al. in prep), which is the only known jetted TDE discovered in the optical \citep{Andreoni2022}. The initial rapid decay is captured in LS4 FOOT survey observations and missed by LSST, which nevertheless continues to detect the transient for several months after it fades below the LS4 detection limit. \textit{Bottom}: Simulation of a low-luminosity TDE that lacks a jet at $z=0.027$, based on the optical evolution of AT\,2020wey \citep{Yao2023}. LSST does not observe the peak of the TDE, but plays an essential role in characterizing the shape and duration of the rising portion of the light curve. For clarity LS4 upper limits are not shown in the bottom panel of this figure. In both the top and bottom panel, symbols and markers are the same as Figure~\ref{fig:synergy}.
    \label{fig:TDE_lc}}
\end{figure*}

High-cadence $i$- and $z$-band observations from LS4 offer a unique opportunity to identify jetted TDEs by sampling the early-time synchrotron spectrum (see Figure~\ref{fig:TDE_lc}, upper panel). Following discovery, LSST will monitor the much fainter long-lasting thermal component. For non-jetted TDEs, LSST can provide very early detections (see Figure~\ref{fig:TDE_lc}, lower panel), to constrain the photometric rise. 

\subsection{Using TDEs to Probe MBH Demographics} \label{subsubsec:tde_statistics}

TDE population studies can address fundamental open questions about MBH demographics, such as their origin. The formation of the very first black holes in the Universe remains unclear. Leading hypotheses suggest that they formed either through the core collapse of Population III stars, which produce black holes with masses of $\sim 10^2\,M_\odot$ (light seeds), or through the direct collapse of gas clouds, which form more massive black holes around $\sim 10^5\,M_\odot$ (heavy seeds). Local dwarf galaxies, with their relatively quiet star formation and merger histories, exhibit similarities to high-redshift galaxies. Therefore, the fraction of local dwarf galaxies that host central black holes --- referred to as the occupation fraction, $f_{\rm occ}$ --- provides insights into early black hole seeding mechanisms. Light seed models, for example, predict higher values of $f_{\rm occ}$ \citep{Greene2020}.

The volumetric rate of TDEs in dwarf galaxies offers a unique way to estimate $f_{\rm occ}$. However, past optical sky surveys have been limited in detecting TDEs in these galaxies because (1) most TDE experiments select transients in known galaxy centers, but dwarf galaxies are highly incomplete in catalogs, and (2) TDEs powered by lower-mass black holes might be intrinsically fainter in the optical band. LS4+LSST can address these challenges. Deep LSST imaging will detect very low-mass host galaxies ($\sim$10$^{7}\,M_\odot$) and monitor TDEs long after they fade below the LS4 detection limit. LS4 will provide high-cadence data during the peak of the light curves, facilitating the selection of TDE candidates.

\subsection{Physics of TDE Optical Emission}

The UV and optical emission from TDEs imply emitting regions that are much larger and cooler than expected for accretion disk emission at the tidal radius. Either X-rays from the accretion disk are being reprocessed by outflows \citep{dai18, Lu20_reprocessing} or an alternative model, such as shocks produced in the intersection of the debris stream \citep{piran15, Huang2024_tde_shock}, is needed to explain the observed emission.

While it is widely accepted that TDE light curves encode the properties of the disrupted star and the mass of the disrupting black hole --- parameters that define physical dimensions and timescales for the TDE --- current light curve fitting methods rely on assumptions about the power sources and emission processes that are still uncertain. For instance, black hole mass estimates derived from light curve fitting sometimes conflict with values inferred from galaxy scaling relations (e.g., \citealt{Hammerstein2023}); and some events exhibit multiple distinct peaks in the optical light curves that cannot be explained with a single process / energy ejection \citep{Ho2025}. Furthermore, some TDEs show broad H, some show broad He, some show both, and some show neither \citep{van-Velzen21, Hammerstein2023}. The origin of these features and the physical distinction between these classes are not yet well understood.

Newly discovered TDEs with LS4$+$LSST will allow us to combine optical light curves with spectroscopic and multi-wavelength follow-up, furthering our efforts to study TDE physics. 

\subsection{Early light curves and Late-time plateaus}

Early bumps (e.g., AT\,2023lli, AT\,2024kmq \citealt{Huang2024_2023lli, Ho2025}), blips, and changes in slope in the rising phase (e.g., AT\,2019azh, AT\,2020zso \citealt{Guo2025_tde_reverberation}) of optical TDEs encode the outflow physics and suggest more than one dominant emission mechanism contributes to the pre-peak light curve. High-cadence LS4 observations will be critical for constraining TDE emission models, ultimately enabling TDEs to serve as a reliable probe of MBH properties.

Several recent studies have shown that the UV light curves for a large fraction of TDEs flatten at late times \citep{vanvelzen19_late_time_UV, Mummery2024_fundamental_scaling_relation}.
The plateau originates as the viscous spreading of the disk competes with the cooling of the disk as the accretion rate declines. Fitting these late-time observations to relativistic accretion disk models shows that the plateau luminosity correlates with the central black hole mass, as estimated from host galaxy properties. LSST will detect these late-time plateaus for LS4-discovered TDEs, providing an independent probe to measure the mass of the central MBH.

\subsection{TDE host galaxies}\label{subsubsec:tde_rate}

Theoretical work has suggested that TDE rates are enhanced in galaxies with asymmetric stellar distributions, which produce torques that drive stars towards the nucleus \citep{Stone2020}. Whether such predictions are borne out is difficult to determine: TDE samples are complicated by the confluence of diverse host galaxy stellar populations and dynamics, unknown levels of dust obscuration, and the unknown role of pre-existing nuclear activity \citep{French2020}.

Current observational work has shown that TDE rates may also be enhanced in galaxies with post-starburst, or E$+$A, stellar populations \citep{Arcavi14, french16, Law-Smith17, Hammerstein21}. The precise reason for this enhancement is not known, but the possible post-merger nature of these galaxies may be responsible for the boost in the TDE rate \citep{stone16, Stone_vV16, Hammerstein21, Hammerstein23_IFU}. Uncovering the true enhancement of these galaxies in TDE populations is key to understanding the mechanisms by which TDEs are triggered, and this can only be done with a larger sample of TDEs. As described in Section~\ref{subsec:scene_modeling}, deep LSST imaging will enable detailed studies of the host morphologies and spectral energy distributions (SEDs) of LS4 TDEs.

\subsection{Irregular (and Regular) AGN variability}

LS4 and LSST will improve the time-sampling and spectral coverage of AGN to probe variations in accretion or jet activity with unprecedented precision \citep[e.g.,][]{LSST_ScienceBook09}. This census will help define ``outliers'', such as rare changing-look AGN (CLAGN; e.g., \citealt{macleod16,ricci23}), turn-on activity \citep{gezari17,nyland20, SanchezSaez24} and narrow-line Seyfert 1-related transients \citep{nlsy1}. An important emerging result is that the disks in these variable AGN appear to change accretion modes faster than would be expected from the thin-disk viscous timescale. The LS4 LEG and FOOT surveys will make key progress in understanding these rarities by discovering them on the rise.

Periodic variations in AGN light curves can reveal binary MBHs situated at the center of AGN (e.g., \citealt{graham15b,graham15a_crts,binary_ptf,binary_panstarrs,Li23}). High-resolution Very Long Baseline Interferometry (VLBI) can spatially resolve only low-redshift binary AGN (e.g., \citealt{rodriguez06}) or larger separation dual AGN (e.g., \citealt{veres21}), whereas long-term photometric monitoring can reveal small separation AGN binaries. The multi-year duration of LS4 will enable the search for long-term periodicities in AGN light curves.

\subsection{Other Nuclear Transients}
Galactic nuclei also play host to other time-variable phenomena, including extreme AGN flares and nuclear supernovae. In particular, ``Ambiguous Nuclear Transients" \cite[ANTs;][]{hol21} and ``Bowen Fluorescence Flares" \cite[BFFs;][]{trakh19} have blurred the line between TDEs and extreme AGN accretion episodes, requiring larger samples to determine their origins.

ANTs exhibit properties of both TDEs and AGN, with some occurring in previously inactive galaxies. Their light curves resemble TDEs but evolve more slowly, last hundreds of days, and often reach higher luminosities ($L_{\rm peak}\gtrsim10^{44}$\,erg\,s$^{-1}$), although some are fainter (e.g. AT\,2018zf and AT\,2020ohl, \citealt{tra19,ric20,lah22,hin22a,hin22b}). Their radiated energy $10^{51} \lesssim E\lesssim 10^{53}$ erg \citep{subrayan23, wiseman23, wiseman25} makes some of them (so-called ``Extreme Nuclear Transients'' or ENTs) the most energetic long-lived transients in the Universe. 
Many ANTs show mid-IR flares and UV/optical/IR emission consistent with dust sublimation, suggesting reprocessed light from a dusty torus \citep{jia17,pet23, oates24,hinkle24}. LS4 will efficiently detect new ANTs, enabling studies of whether they arise from novel AGN accretion modes, massive TDEs, or spinning MBHs. Its $z$-band sensitivity will also track mid-IR flares, offering insights into ANT emission mechanisms.

BFFs are a recently discovered class of accreting MBHs, characterized by steep-rise, slow-decline optical flares, bright UV and radio emission, and AGN-like spectra with emission lines driven by Bowen fluorescence \citep{bowen}. 
Their flares persist for $\sim$1\,yr, sometimes with secondary flares, and may represent TDEs occurring within AGN disks \citep{Veres24TDEneutrino}. High-cadence LS4 observations will capture BFFs on the rise, while combined LS4+LSST data will track their decline and color evolution, distinguishing them from standard AGN variability.

\subsection{Classification and follow-up efficiency} \label{subsubsec:mbh_follow_up}

Low-level AGN variability detected by LSST can rule out ``normal'' AGN activity and reduce contamination in samples of TDEs and other unusual accretion events found by LS4. To spectroscopically classify TDEs and other nuclear transients we will use the ESO 4m Multi-Object Spectroscopic Telescope \citep[4MOST;][]{de-Jong19} multi-object spectrograph, which will classify $>10^{5}$ transients (including nuclear ones) as part of the Chilean AGN/Galaxy Extragalactic Survey \citep[ChANGES][]{Bauer2023} and Time Domain Extragalactic Survey \citep[TiDES;][]{Frohmaier2025a} campaigns, Son of X-Shooter \citep[SoXS;][]{SOXS18}, SDSS-V \citep{Kollmeier2019}, the Dark Energy Spectroscopic Instrument \citep[DESI;][]{DESI-Collaboration16}, and Las Cumbres Observatory \citep[LCO;][]{lco}, where dedicated programs are already in place.

\section{Physics of Stellar Explosions}

The vast majority of extragalactic transients found by LS4 will be SNe, which will provide a significant opportunity to advance our understanding of the explosions that punctuate the end of some star's lives. In the past decade, PTF, Pan-STARRS, and ZTF have demonstrated that a combination of very early, multi-color detections and long observational baselines, both before and after the explosion, are critical to unravel the nature of SNe. As noted in Section~\ref{sec:lsst-synergy}, LS4$+$LSST will build on this legacy while uniquely probing the years before and after explosion to unprecedented depths for thousands of SNe. This will provide the global SN community ample time to trigger critical follow-up observations, including high-resolution spectroscopy, infrared imaging and spectroscopy, spectropolarimetry, and X-ray and radio follow-up. Below we highlight some of the most interesting science cases that can be pursued with LS4 alone, as well as the synergistic power of combining LS4 with LSST.

\subsection{Infant Core-collapse Supernovae}\label{sec:infant_ccsne}

The LS4 FOOT survey (see Section~\ref{sec:survey:foot}) is designed to routinely discover SNe within 48\,hr of first light. Such discoveries require the rapid vetting of candidates and follow-up. Latencies in vetting and follow-up remain a major bottleneck in infant transient studies, but automation can reduce the time lag. In its alert stream, LS4 will employ image-based deep learning for young transient identification. 
Building from the success of \texttt{BTSbot} \citep{Rehemtulla2024a}, which identifies ZTF alerts that originate from bright ($m_{\rm peak} < 18.5$ mag) extragalactic transients, LS4 will integrate alert streams from multiple surveys to increase the rate of these discoveries. 
By incorporating autonomous transient identification and follow-up tools \citep[e.g.,][]{Rehemtulla2025a}, LS4 will pursue similar no-human-in-the-loop automation, reducing follow-up latency from next night spectroscopy to minutes or hours. 

Spectra of 1--2\,d old SNe guarantee a rich yield of transient emission and absorption features. Numerous studies have shown that a large fraction of core-collapse supernovae (CCSNe) are engulfed in compact ($\sim$10$^{14}$\,cm) shells of circumstellar material (CSM)  that is swept up by the SN ejecta within days of explosion \citep[e.g.,][]{Bruch21, Bruch23,Jacobson-Galan24}. In response to the radiation field of the SN shock break-out, this CSM is excited and ionized, producing a series of recombination lines that can be visible for several days. As demonstrated by the analysis of ``flash spectroscopy'' data \citep{Niemela1985a, AGY+14, Yaron17}, the spectral lines that appear following the excitation and ionization of the CSM encode information of the surface composition of the dying star just before it explodes \citep{Groh2014a, Gal-Yam22, Schulze24}, the mass-loss history of the progenitor star \citep{Yaron17, Boian2020a, Jacobson-Galan24}, and details about the SN shock propagation \citep{Zimmerman2024a}. Together, these data enhance our understanding of what massive stars do just before they die and the SN explosion mechanisms themselves. Furthermore, combined with pre-SN imaging from LSST, reaching 2--3 magnitudes deeper than previous surveys, this method can constrain the progenitor's mass-loss history over timescales of years, months, and weeks before the explosion \citep{Ofek14, Strotjohann2021a, Jacobson-Galan22}, offering an unprecedented window into the final evolutionary stages of massive stars.

\subsection{The Mass-loss History of Dying Massive Stars}

Hydrogen-poor CCSNe arise from massive stars (zero-age main sequence mass $\gtrsim8~M_\odot$) that lose most or all their hydrogen envelopes before core collapse. These SNe, collectively known as stripped-enveloped SNe (SESNe), include several subtypes \citep{Filippenko97, Gal-Yam17}: Type IIb SNe (which retain some hydrogen), Type Ib SNe (which lack hydrogen but contain some helium) and Type Ic SNe (which have neither hydrogen nor helium). Extensive studies of SESNe suggest that their low ejecta masses are inconsistent with their mass loss being driven solely by stellar winds \citep[e.g.,][]{Taddia15, Prentice19}, which favors a binary origin instead. However, if these SNe indeed result from binary interactions, they should be surrounded by a significant amount of CSM, with its mass and distribution likely influenced by the binary separation \citep{2014smith}.

Some SESNe exhibit multiple peaks in their early light curves, likely caused by the shock breakout at either the surface of the progenitor or through an extended envelope surrounding it \citep{Woosley94a, Bersten12, Piro15}. Large sample studies \citep[e.g.][]{Das24} show that these events contribute between 3 and 10\% of the SESN rate. The LS4 FOOT survey (Section~\ref{sec:survey:foot}) will allow us to measure the rate and properties of initial peaks for a large number of SN classes (e.g., IIb, Ib, Ic, Ic-BL) to improve our  understanding of what massive stars do just before they die.

In recent years, an increasing number of SESNe have shown double-peaked light curves at later stages, likely caused by the interaction of the SN ejecta with CSM. In some cases, the SNe simultaneously show a spectroscopic metamorphosis into a CSM-powered SN (e.g., SN~2014C, \citealt{Milisavljevic15}; SN~2017ens, \citealt{Chen2018a}), whereas in others such a spectroscopic transformation is either subtle (SN~2022xxf, \citealt{Kuncarayakti23}; SN 2022jli, \citealt{Chen2023b, Moore23}) or even absent (e.g., SN~2019tsf, \citealt{Sollerman20}; SN~2019cad, \citealt{Gutierrez21}; SN~2023aew, \citealt{Kangas24,Sharma24}) despite a clear increase in optical flux. The reasons for that are debated. Combining LS4 with LSST will allow us to monitor SN light curves for hundreds of days and, therefore, not only measure the rate and properties of rebrightenings but also how the rebrightenings evolve as a function of the light curve phase and SN type. Acquiring well-timed spectra provides the unique opportunity to gain a much-improved understanding of the physics of SN ejecta$+$CSM interaction, the pre-SN mass-loss episodes (stellar winds vs.\ episodic outburst vs.\ binary-driven envelope stripping; \citealt{2014smith}, \citealt{2018morozova}), and ultimately about the uncharted late-stage evolution of dying massive stars.

Many interacting SNe exhibit rebrightenings, undulations and irregular declines \citep{2020nyholm}, indicating the presence of clumpy, multi-shell, or asymmetric CSM structures. The LS4 FOOT survey will characterize these fluctuations from geometric asymmetries and density variations \citep{2012Kiewe,Ofek14}, differentiating between steady mass loss and episodic/explosive ejections and uncovering the physics governing the CSM interaction processes \citep{Moriya2013a, Ofek14, Khatami2024a}. Long-term monitoring, which will be provided by LS4$+$LSST, will uncover additional events like iPTF14hls \citep{Arcavi14hls2017Natur.551..210A}, which had a long-lived undulating light curve that was suggested to be powered by magnetars, fall-back accretion, or CSM interaction. Only with observations collected several years after explosion would some of these scenarios be ruled out 
\citep{Sollerman14hls2019A&A...621A..30S}.

\begin{figure}
\centering
\includegraphics[width=3.2in]{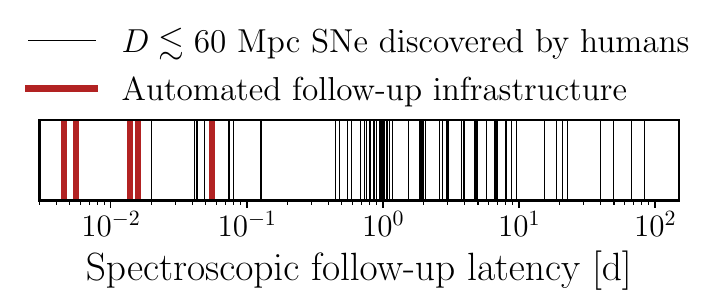}
\caption{
Distribution of spectroscopic follow-up latencies for automated and human-in-the-loop workflows in ZTF; figure reproduced from \cite{Rehemtulla2025a} and \cite{Rehemtulla2025b}. Latency is defined as the time elapsed between static alert filter being passed and the first spectrum being obtained. Automating transient identification and follow-up significantly expedites observations of valuable nearby SNe, enabling more detailed investigations of early-time phenomena like the flash features seen in CCSNe.
}
\end{figure}

 \subsection{Engine-driven transients} \label{SubSec:engine}
Engine-driven astronomical transients, a heterogeneous class, are powered by black holes (BHs) and neutron stars (NSs). Examples include the recently discovered Fast Blue Optical Transients (FBOTs, e.g., \citealt{Drout14,Ho22}) and gamma-ray bursts (GRBs), which come from massive stellar explosions and binary NS mergers.
Engine-driven transients can launch relativistic outflows with a range of collimation properties, which is a consequence of the deep gravitational potentials of their compact objects. As a result, the electromagnetic evolution of these events evolves on shorter timescales than normal SNe. The LS4 FOOT survey is designed to capture the $\sim$1\,d timescales associated with engine-driven events. 

A decade after the discovery of FBOTs \citep{Drout14}, it is now clear that low-luminosity FBOTs represent the ``tail'' of already known phenomena (e.g., cooling-envelope emission from Type IIb SNe), and that Luminous FBOTs (LFBOTs hereafter) represent a physically distinct class of events (e.g., \citealt{Ho22}). Reaching a peak luminosity $L_\mathrm{peak}\approx 10^{45}\,\mathrm{erg\,s^{-1}}$ in a few days, LFBOTs challenge models invoked to explain ordinary stellar explosions and require more exotic scenarios. While the intrinsic nature of LFBOTs is still highly debated (with interpretations ranging from jetted stellar explosions to TDEs on Intermediate Mass Black Holes; IMBHs), X-ray and radio observations of a handful of LFBOTs have led to the discovery of a rich phenomenology, including mildly relativistic ejecta \citep[e.g., CSS\,161010,][]{Coppejans20}. 

At the same time, the red sensitivity of LS4 is well matched to GRB afterglows, which are powered by synchrotron emission. In this way LS4 can play a critical role in opening to discovery the phase space of ``orphan afterglows'' in the local Universe (i.e., GRB afterglows that eluded high-energy telescopes either because of satellite sensitivity or pointing at the time of the GRB), as well as  GRB events for which the jet is intrinsically absent (e.g., choked by baryons like in baryon-loaded fireballs; \citealt{Cenko12}). 

\subsection{Superluminous Supernovae}
Hydrogen-poor superluminous supernovae \citep[SLSNe-I;][]{Quimby2011a} are a rare ($5.6^{+5.4}_{-2.8}~\rm Gpc^{-3}\,yr^{-1}$; \citealt{Perley2020a}) class of transients with optical luminosities between $-23$ and $-20$\,mag at peak, about a factor of 10--100 more luminous than ordinary core-collapse supernovae \citep{DeCia2018a, Lunnan2018b, Angus2019a, Chen2023a}. Despite recent observational breakthroughs \citep[e.g.,][]{Nicholl2015a, Nicholl2016b, Yan2017a, Yan2017b, Bose2018a, Lunnan2018a, Quimby2018a}, the underlying physical mechanisms that drive their extreme luminosities remain debated \citep{Moriya2018a, Gal-Yam2019a}. They are thought to be powered by one of the following: the spin-down of a rapidly rotating young magnetar \citep[][]{Kasen2010a}, interaction of the SN ejecta with a massive (3--5 $M_\odot$) C/O-rich CSM \citep[][]{Blinnikov2010a, Baklanov2015a}, pulsational pair-instabilities in which collisions between high-velocity shells are the source of multiple, bright optical transients \citep[PPISNe;][]{Woosley2007a}, or tidal disruption of the companion star by a compact object born in a binary system \citep{tsuna25_SLSNe}.

Light curves are essential to understand the immense source of energy and progenitors of SLSNe. The ZTF 3-d cadence Northern Sky Variability Survey \citep{Bellm19a} revealed photometric bumps and wiggles in a large number of SLSNe-I \citep[][see also \citealt{Hosseinzadeh2022a}]{Chen2023b}, likely due to CSM interaction. The LS4 LEG Survey (Section \ref{sec:survey:leg}) will capture photometric bumps and wiggles in the $giz$ light curves of many SLSNe-I ($z \lesssim 0.2$). 

WFD will supplement LS4-discovered SLSNe by capturing bumps on the rising portion of the light curve (luminosity: $-16$ to $-20$~mag; duration: $\lesssim 20$--30\,d; \citealt{Leloudas2012a, Nicholl2015a, Nicholl2016a, Smith2016a, Angus2019a}) or slowly rising plateaus before the ``normal'' rise \citep{Anderson2018a}. Furthermore, while LS4 probes the photospheric phase, LSST will monitor SLSNe-I for 100s of days into the nebular phase. These late-time observations are very sensitive to the SLSN powering mechanism including magnetars and radioactive decay. Finally, deep LSST observations will be especially useful for identifying the faint, star-forming dwarf galaxies associated with SLSNe-I \citep{Lunnan2014a, Leloudas2015a, Perley2016a, Chen2017a, Schulze2018a}. Detailed knowledge of the host galaxies has the potential to transform out understanding of the powering mechanisms, progenitors, and physics that operate within SLSNe. 

LSST will discover thousands of SLSNe, but the sparse cadence in WFD is insufficient to fit models and infer critical parameters like the ejected mass. \cite{2018ApJ...869..166V} found that they could accurately recover the parameters of SLSN models injected into simulated LSST data only 18\% of the time. LS4 will dramatically improve this situation by resolving the light-curve peaks for hundreds of $z \lesssim 0.5$ SLSNe, allowing accurate measurements of the ejecta mass distribution to constrain SLSN progenitor models.

\subsection{Early Observations of Type Ia Supernovae}\label{Subsubsec:earlyIa}

Observations of SNe\,Ia shortly after explosion can strongly constrain theoretical predictions about the explosion mechanism. Resolved ``bumps'' in the early optical emission of several SNe\,Ia have been detected \citep[e.g.,][]{Goobar15, Jiang17, Hosseinzadeh17,Shappee19,Dimitriadis19, Deckers22,Srivastav23}. In addition, strong, early UV emission has now been detected in two SNe, iPTF\,14atg \citep{Cao15} and SN\,2019yvq \citep{2020ApJ...898...56M, Tucker21,Burke2021a}, though their subsequent evolution was very different. Multiple models could explain the early excess flux, including shock interaction with a non-degenerate companion star \citep{Kasen10a}, helium shell burning in the double-detonation of a sub-Chandrasekhar mass white dwarf \citep{Shen18a,Polin19}, CSM interaction \citep{Piro16}, or the presence of radioactive $^{56}$Ni in the outer ejecta \citep{Piro12,Magee20a}. The LS4 FOOT survey combined with advances in the autonomous identification of infant SNe will allow us to routinely discover infant Type Ia SNe and swiftly obtain the critical follow-up observations (see also Section \ref{sec:infant_ccsne}).

\subsection{Intrinsically Red and Dust-obscured Stellar Explosions}

The frequency and depth of LS4 $i$- and $z$-band observations provide a unique opportunity to search for exceptionally red stellar explosions, an area that is historically underexplored.
Recent discoveries demonstrate that some exotic explosive events have intrinsically red optical colors ($g-i\gtrsim1$\,mag), including a new class of peculiar SNe\,Ia thought to be associated with sub-Chandrasekhar mass white dwarf (WD) progenitors \citep{Jiang17,De19,Liu23} as well as a variety of underluminous ``gap'' transients \citep{munari02, 2011tylenda}. Red-optical colors are also observed in normal SNe (especially some SESNe), often due to extrinsic factors such as line-of-sight extinction from their dusty stellar nurseries. LS4 will be proficient in hunting these red transients, which are systematically missed in surveys optimized for the blue-optical.

Extreme line blanketing from iron-group elements (IGEs) in the maximum-light spectra of SNe\,Ia provide strong evidence for the helium-shell double-detonation explosion of a sub-Chandrasekhar mass WD \citep[e.g.,][]{Polin19}. After a brief flash powered by radioactive decay, the IGEs formed in the helium-shell detonation will obscure most of the UV and blue-optical flux from the SN, leading to a dramatic blue--red color inversion on the rise and an exceptionally red color at peak \citep[e.g.,][]{Noebauer17,Polin19}. Several candidate double-detonation explosions have been found \citep[e.g.,][]{Jiang17,De19,Dong22,Padilla-Gonzalez23,Liu23}, and the red-sensitivity of LS4 will readily identify the pronounced flux suppression seen in these SNe.

While CCSNe are usually extremely luminous, some massive stars produce red transients that fall in the nova-SN luminosity ``gap'' \citep{kasliwal11}, such as luminous red novae (LRNe) and intermediate luminosity red transients (ILRTs).
LRNe are usually associated with binary stars that have undergone the common-envelope phase \citep[e.g.,][]{2011tylenda, Nandez14, Chen24a}. While most LRNe are intrinsically faint ($M_V\gtrsim-6$\,mag) events in the Milky Way, there is a growing sample of bright ($M_V\lesssim-12$\,mag), extragalactic LRNe \citep{Pastorello21}. ILRTs exhibit red continuua with Balmer emission lines and, in some cases, extremely long-lived outbursts \citep{smith09}. They remain enigmatic with evidence pointing to a variety of potential origins, including electron-capture induced collapse from extreme asymptotic giant branch stars \citep[e.g.,][]{botticella09} and outbursts from luminous blue variable-like stars \citep[e.g.,][]{Jencson19}. The systematic identification of LRNe and ILRTs will provide insight into rare and understudied phases of stellar evolution.

LS4 may also discover an entirely new population of deeply enshrouded SNe, providing a more complete picture of massive star explosions. Many SNe, particularly those in dense, metal-rich star-forming regions, experience significant circumstellar dust reprocessing, leading to optical suppression \citep{2016bocchio, 2025serrano}. Finding these SNe is essential for constraining intrinsic transient rates, where it is known that many SNe are missed due to line-of-sight extinction \citep{Kasliwal17a,Jencson19}.  This is especially pressing in galaxies with the highest star-formation rates as their nuclear regions exhibit both the youngest stellar populations and largest line-of-sight extinction \citep[e.g.,][]{Richmond1998}.  LS4 will be significantly more sensitive to these transients due to the sensitive and regular observations in the $i$- and $z$-bands.

\subsection{Calcium-strong transients}

Relative to normal SNe, Calcium-strong (Ca-strong) transients are intrinsically faint and fast-evolving SNe with uncertain origins \citep[e.g.,][]{Kasliwal2012, milisavljevic17, shen19, De20, Jacobson-Galan2020}. The photospheric phase spectra of Ca-strong transients are dominated by helium features and resemble SNe~Ib \citep{Perets10}, while lacking the IGEs typically seen in thermonuclear SNe. Their spectra rapidly become optically thin, and their nebular phase spectra exhibit strong [\ion{Ca}{2}] emission that dominates over relatively weak or non-existent [\ion{O}{1}] \citep{2020ApJ...905...58D}. They are often found in remote environments, far from candidate elliptical/S0 host galaxies \citep{Kasliwal2012,2013MNRAS.432.1680Y,2017ApJ...836...60L}. Deep observational limits at the location of Ca-strong candidates show no signs of star formation or globular cluster hosts \citep{2014MNRAS.444.2157L,2017ApJ...836...60L}. Thus, while they spectroscopically resemble SNe~Ib with massive star progenitors, their preference for old stellar populations and remote environments favors a white dwarf progenitor \citep{shen19}. 

The underlying explosion mechanism for these events is still hotly contested. On the massive star side, they could arise from the lowest mass (8--12~$M_\odot$) stars as an ``electron-capture supernova'' \citep{Nomoto1984, Kawabata2010}, or ultra-stripped stars \citep{Tauris2013, Tauris2015, Moriya2016}. White dwarf origin channels have been suggested to include the merger of a C/O WD with a hybrid He/C/O WD \citep[e.g.,][]{Jacobson-Galan2020, 2023ApJ...944...22Z}, helium-shell detonations on low-mass CO WD \citep[e.g.,][]{2011ApJ...738...21W,2019ApJ...873...84P,2023ApJ...944...22Z}, WD-NS mergers \citep{2018A&A...619A..53T,2019MNRAS.486.1805Z}, and WD mergers with IMBHs \citep{2015MNRAS.450.4198S}. More speculative scenarios have been proposed to explain their large offset from galaxy centers involving a double WD system ejected from its host galaxy by interaction with a super-massive BH (SMBH; \citealt{Foley2015}), or single WD explosions, triggered by traversing asteroid-mass primordial BHs in dwarf satellite galaxies \citep{2024PhRvL.132o1401S}.

Early photometric observations of some objects have revealed double-peaked light curves, where the first peak is short-lived and associated with X-rays, likely indicating CSM interaction \citep[e.g.][but see also \citealt{2007ApJ...662L..95B} where decay of short-lived isotopes such as $^{48}$Cr can result in early flux excess]{Jacobson-Galan2020, Jacobson-Galan2022}. The LS4 FOOT survey will capture early initial peaks and flux excesses from Ca-strong transients, while the red-sensitivity will enable long-term monitoring as the flux rapidly fades in the $g$ band. High-resolution images from LSST will place deep limits on the explosion environment of all LS4-discovered Ca-strong events.

\subsection{4MOST-TiDES Spectroscopic Follow-up}

4MOST is a fiber-fed spectroscopic facility that packs ${\sim}$2,400 configurable science fibers into a 4.2\,deg$^{2}$ FOV \citep[][]{de-Jong19} . Over its 5\,yr duration 4MOST TiDES \citep[][]{Frohmaier2025a} will use 250,000 fiber-hours to obtain spectroscopic observations of transients and their host galaxies. The strategy for TiDES is simple, wherever 4MOST points within its $\sim$18,000\,$\mathrm{deg}^2$ footprint, any live-and-bright ($r\leq22.5$\,mag) transients discovered by feeder surveys such as LS4 and LSST will have fibers placed on them. The host galaxies of transients fainter than the TiDES limit will obtain host-galaxy spectra. Every LS4-discovered transient will be brighter than the 4MOST observational limit (for a typical 2,700\,s exposure), meaning all LS4 SNe will be added into the TiDES observational queue.

\subsection{Gravitationally-lensed Supernovae}

glSNe are a new and novel probe of cosmology and high-redshift astrophysics. The multiple images of glSNe produced by strong lensing permit a direct measurement of $H_0$ that is independent from the cosmic distance ladder, while magnification enables intrinsically fainter and more distant sources to be discoverable. Current estimates predict that $H_0$ can be measured to 1.5\% with 3-years of observations with Rubin \citep{Arendse24}. glSN discoveries to date have mostly been serendipitous from deep, high-resolution images \citep{Kelly15,Rodney21,Kelly22,Chen22,Frye24,Pierel24}, with only three confirmed discoveries by wide-field surveys to-date \citep{Quimby13,Goobar17,Goobar23}. In the next decade, the unprecedented volume probed by LSST will enable the first statistical sample of glSNe, with estimates reaching up to $\sim$100\,yr$^{-1}$  \citep{OM10,Goldstein2019,SainzDeMurieta23,Arendse24}. LS4 will support the discovery and science possible with glSNe.

Wide-field surveys that monitor known lensed galaxies can efficiently discover glSNe. This can be achieved with a ``watchlist'' of known lenses, and can also include candidate lenses, which typically consist of luminous red galaxies (LRGs) and galaxy groups and clusters, which collectively make up a large proportion of the total strong lensing cross-section in the Universe \citep[e.g.][]{Robertson2020}. Discoveries by observations of resolved multiple images will also be possible for glSNe in the upper tails of the image separation distribution: $\sim5$\% will have image separations larger than the LS4 pixel scale, which can be resolved either directly or via PSF deconvolution \citep{SainzDeMurieta23,Millon24}.

Due to their standardizable nature, glSNe\,Ia can be identified when a SN\,Ia is far too luminous for the redshift of its ``host,'' which in the case of a glSN found in ground-based imaging is the lens galaxy rather than the true SN host galaxy \citep[e.g.,][]{Goobar17, Goobar23}. Spectroscopic observations were essential to identify the glSNe found by ground-based surveys, but most glSN candidates will be faint and not receive a spectrum. With a higher cadence than LSST WFD, LS4 light curves will enable the photometric classification of SNe\,Ia. Furthermore, the red-sensitivity of LS4 will help to identify glSNe which will occur at substantially higher redshifts than the ``normal'' unlensed LS4 SN\,Ia population.

Assuming an $i$-band detection limit of $m_i = 20.5$~mag, a $\sim$half-sky survey should discover $\simeq$0.2\;glSNe\;yr$^{-1}$ \citep{SainzDeMurieta23}. While this estimate is somewhat small, the true number of known glSNe observed with LS4 will be larger than this in reality, since LSST-discovered systems will permit sub-threshold measurements of glSNe within LS4. Finally, to increase the discovery rate of glSNe we will perform offline stacking of LS4 LEG and FOOT images to search for slow-evolving, red SNe below the LS4 detection threshold. While these candidates will be faint, they will still be brighter than the detection limit of 4MOST/TiDES and SoXS, which will enable follow-up of these highly valuable targets.


\section{Multi-messenger Astrophysics}

\subsection{Electromagnetic (EM) Follow-up of Gravitational Wave sources and poorly localized GRBs with LS4 and LSST}
\label{SubSubSec:GWs}

The discovery of the an EM counterpart to the binary NS merger, GW\,170817, launched the era of GW$+$EM multi-messenger astrophysics \citep{Abbott17a,MarguttiChornock2021}.  While the direct detection of GWs from astrophysical sources has enabled an exciting new view of the cosmos, GW signals paired with EM observations provide an unprecedented probe of astrophysics. The first EM signal from GW\,170817 was a short-duration GRB (SGRB), GRB\,170817A, which provided direct evidence that at least some SGRBs come from NS-NS mergers \citep{gw170817grb}. As observations across the EM spectrum of GW\,170817 demonstrated, the identification of an EM counterpart provides immense scientific benefits, including: improved localization leading to a solid host-galaxy identification;  determination of the source  distance and energy; characterization of the progenitor's local environment; the ability to break model degeneracies between distance and inclination of the binary system; and insight into the hydrodynamics of the merger. Furthermore, identification of the EM counterpart facilitates other fields of study such as determining the primary sites of heavy $r$-process element production, placing limits on the NS equation of state, making independent measurements of the local Hubble constant, $H_0$, and further elucidating their connection to SGRBs \citep{Abbott17a,AbbottH0,Hjorth17}. See  \cite{MarguttiChornock2021} for a review of the EM+GW properties of GW\,170817 and the immense  scientific impact of that discovery. 

\begin{figure*}
\centering
\includegraphics[width=6.75in]{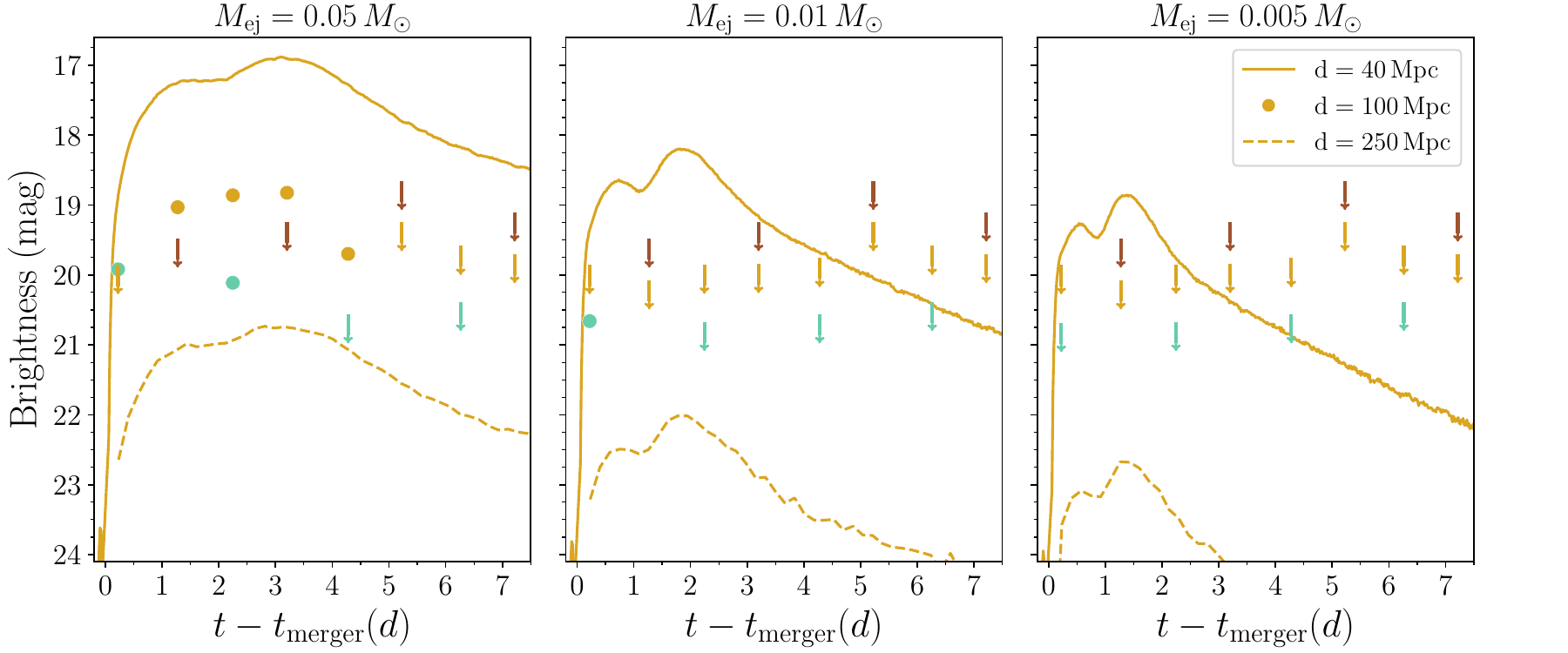}
\caption{Simulated LS4 light curves for a kilonova (KN) with ejecta velocity $v_\mathrm{ej}=0.15\,c$ and different ejecta masses, $M_\mathrm{ej}=0.05, \;0.01, \;0.005\,\rm{M_{\odot}}$, shown in the left, middle, and right panels, respectively, each at three representative distances of 40, 100 and 250\,Mpc. The KN luminosity is not a strong function of $v_\mathrm{ej}$ and values within $0.1$--$0.2\,c$ give comparable results. The $0.05\,\mathrm{M_{\odot}}$ model is closest to the observed KN properties of GW\,170817. The LS4 $g$, $i$, and $z$ light curves for observations starting $\sim$6\,hr after the NS-NS merger are shown for the 100\,Mpc model, while only the simulated $i$-band evolution is shown for the 40 and 250\,Mpc models. The panels extend to $i \approx 24.5$\,mag, which is the typical detection limit for LSST in a 30\,s exposure\citep{SMTN-002}. When the ejecta mass is large, LS4 can detect a KN to distances $\gtrsim$100\,Mpc. For much lower masses, LS4 will only detect KNe to a few 10s of Mpc. Symbols the same as Figure~\ref{fig:synergy}.
\label{Fig:KN2component}}
\end{figure*}

Exploiting the success of multi-messenger astronomy in the next decade will require the continued investment in observational resources. The fourth observing run (O4) of the LVK collaboration is currently planned to end in October 2025.  Upon completion of instrumental upgrades, LIGO, Virgo and KAGRA are expected to re-start operations with a fifth observing run (O5) in 2027 with improved sensitivity and localizations \footnote{\url{https://www.ligo.caltech.edu/page/observing-plans}} resulting in improved capabilities for counterpart discovery \citep{2023ApJ...958..158K,2024MNRAS.528.1109S}. With its 5\,yr duration, LS4 will overlap with the tail end of O4 and all of O5. Furthermore, LSST also begins in the second half of 2025 and up to 4\% of the LSST observing time has been approved for Target of Opportunity (ToO) programs (see \citealt{AndreoniMargutti2024LSSTToO} and references therein). 
LS4, in concert with LSST ToO observations, will provide an agile, wide field-of-view system that is well suited for GW follow-up. Synergy between LS4 and LSST operations will maximize the scientific potential and discovery power of both surveys. For example, the best use of the large aperture of LSST is to follow up the most distant mergers \citep{Margutti18,Andreoni22}, which will have the faintest EM counterparts that are  simply out of reach for any other observing facility (Figure~\ref{Fig:KN2component}). Together, LSST and LS4 can be the premiere discovery engine of EM counterparts to GW sources in the Southern Hemisphere.

Meanwhile, the \textit{Fermi} satellite \citep{mlb+09} is the most prolific workhorse for GRB discovery, with $\approx 240$ new GRBs per year \citep{bmv+16}. Given that the typical GRB localizations are $\sim 50-100$\,deg$^2$ \citep{cbg+16}, few facilities attempt to find Fermi GRB afterglows. Fermi-discovered SGRBs, especially those with characteristics similar to GRB\,170817A, are especially interesting because they may be hiding in the nearby universe \citep{vvr+19}. SGRBs provide a beacon for finding NS mergers in the local Universe (and when GW interferometers are offline this represents the most reliable path to discovery).

\begin{figure*}
\centering
\includegraphics[width=2.7in]{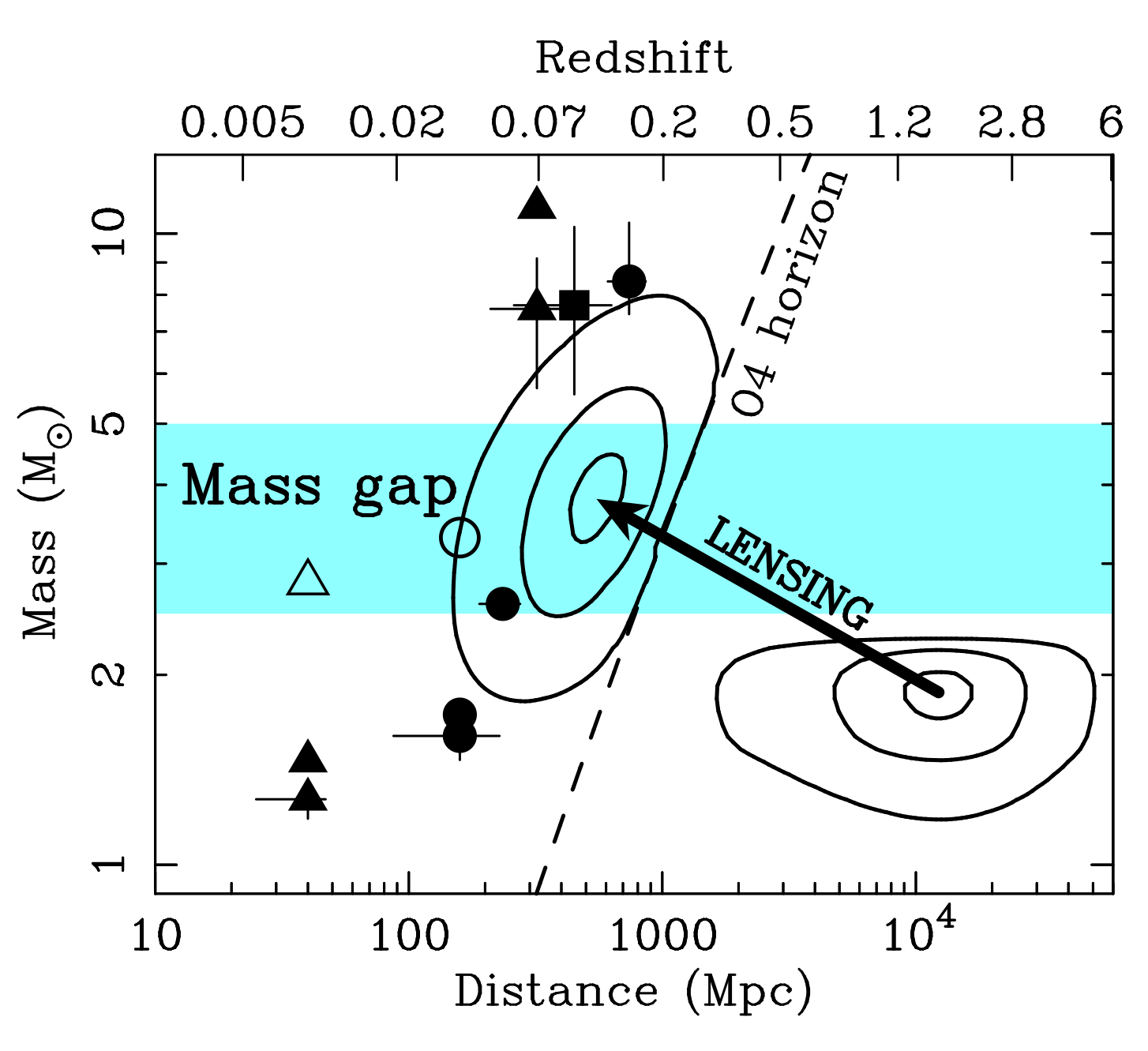}\hspace{5mm}
\includegraphics[width=3.6in]{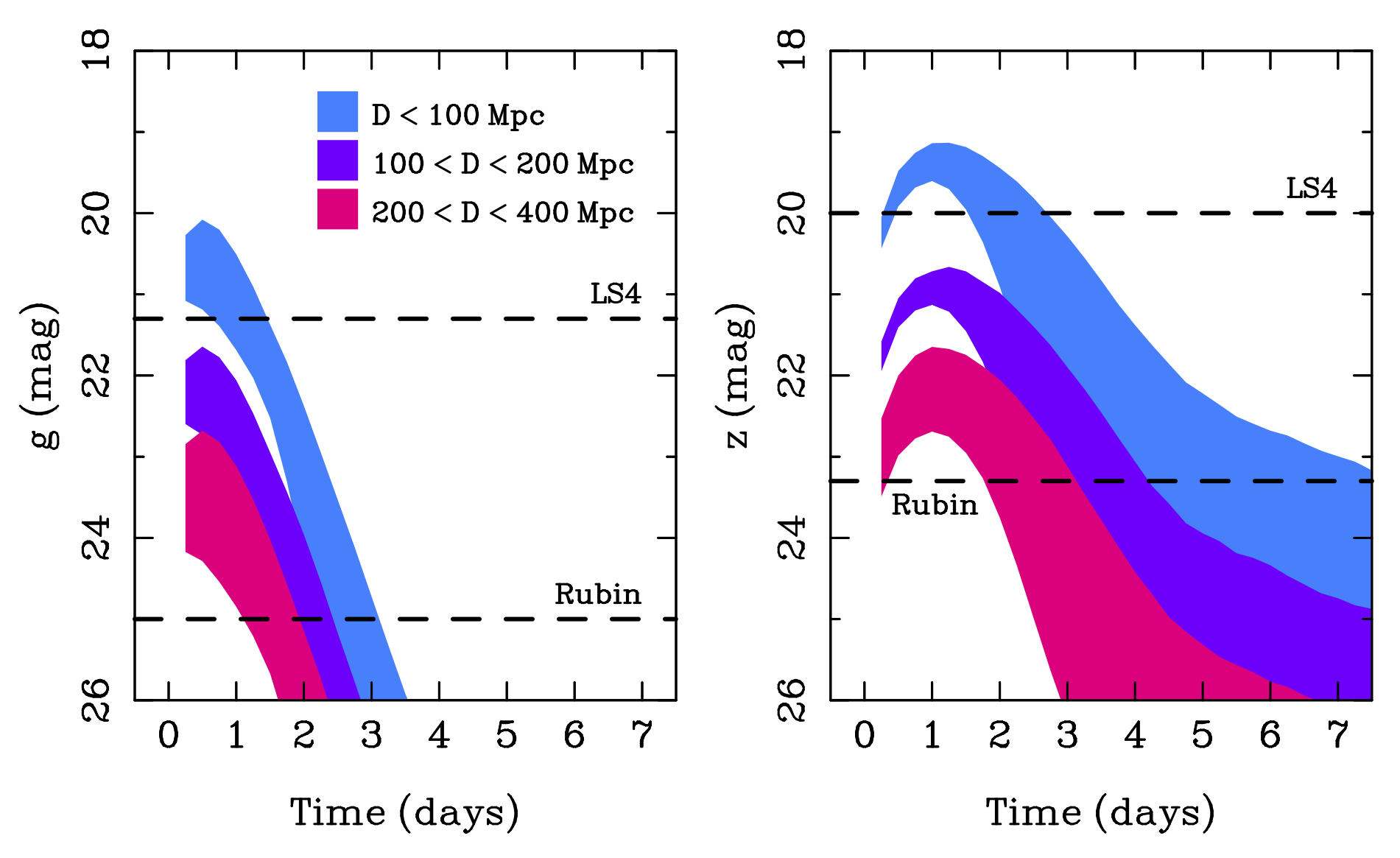}
\caption{\emph{Left}: Mass-distance plane for binary compact object mergers detected by LVK, showing how lensing makes distant binary NS mergers appear as being more massive and closer than they really are. 
We expect most of 
the lensed BNS mergers will be inferred by the LVK in low latency to be located in the mass gap between the heaviest NS and the lightest BHs.  
Contours show the 90th, 50th, and 10th percentile of the predicted lensed population as inferred by LVK assuming no lensing ($\mu=1$) bright-ward of the horizon, and the true distribution faint-ward of the horizon (dashed line).  Filled symbols are the masses of individual compact objects and open symbols are the merger remnant.  Squares, triangles and circles show detections from LVK runs O1, O2, and O3, respectively.   \emph{Centre and Right:} Predicted $g$- and $z$-band light curves of lensed KN counterparts to lensed binary NS mergers, as a function of the distance to them that LVK infer in low latency assuming $\mu=1$. 
LS4 target of of opportunity observations of regular depth will have the sensitivity to detect the brightest (apparently loudest and nearest) lensed events (upper dashed line), with strong potential to reach fainter systems by integrating for longer on the best localized events, following the approach outlined by \cite{AndreoniMargutti2024LSSTToO}. The relative sensitivity of LS4 and Rubin is indicated by the difference between upper and lower dashed lines. The width of the light curve bands show the central 90\% of the predicted probability density for the lensed population. Based on \cite{Smith23LensedGW}. 
\label{Fig:lensedGW}}
\end{figure*}

LS4 will conduct ToO observations of the best-localized GW events and SGRBs in the local Universe (e.g., $d\lesssim 200$\,Mpc),
with these primary scientific goals: 

\begin{itemize}
    \item[(i)] Growing the sample of EM counterparts to GW events to conduct statistically-rigorous systematic studies to understand the diversity of EM emission, their host environments, the nature of merger remnants, and their contribution to the chemical enrichment of the universe through $r$-process production, which shapes the light-curves and colors of the ``kilonovae'' (KNe) associated with GW events (e.g., \citealt{2015MNRAS.446.1115M}).
    
    \item[(ii)] Sampling the very early KN emission (e.g., $\lesssim10$\,hr post-merger) to identify emission mechanisms beyond the KN (e.g., neutron precursor, shock-cooling, e.g., \citealt{PiroKollmeier18}). Despite the fact that the optical counterpart of GW\,170817 was discovered less than 11 hours post-merger (e.g., \citealt{arcavi17,Coulter17,Cowperthwaite17,Drout17,Kasliwal17,Lipunov17,Smartt17,Soares17,Tanvir17,Valenti17,Villar17}), these observations were still unable to definitively determine the nature of the early time emission.
    
    \item[(iii)]Finding the first EM counterpart of a merger of a NS-BH. This system might produce a KN, but the ejecta mass depends on the  mass ratio of the binary and the NS equation of state, and in some cases there may be no material ejected at all (e.g., \citealt{Foucart2018}). It is also unclear if NS-BH mergers will be able to produce the bright early-time blue emission seen in GW\,170817 \citep{2015MNRAS.446.1115M}. 
    
    \item[(iv)] Exploring the unknown of EM counterparts to BH-BH mergers \citep{Graham20} and to unidentified GW sources. Utilizing the LS4 public alerts that fall within the LVK sky maps region we will search for BH-BH GW events.

    \item[(v)] Measuring $H_0$ to 2-8\% precision \citep{2019BAAS...51c.310P} using the standard siren method \citep{1986Natur.323..310S} for the GW events with an identified counterpart. These constraints can further be improved with well-sampled KN light-curve observations \citep{2020ApJ...888...67D}. A precise and accurate measurements of $H_0$ from standard sirens could help clarify whether the observed $H_0$ tension between late \citep{Riess_2019} and early \citep{planck18} time Universe measurements arises from beyond--$\Lambda$CDM physics or unknown systematics (e.g., \citealt{verde}).
    
    \item[(vi)] Pinpointing the origins of nearby SGRBs, independent of their GW signals, by identifying and localizing their counterparts. 
    
    \item[(vii)] Testing the gravitational lensing interpretation of bright GW sources detected in LVK in the putative gap between the heaviest NS and the smallest stellar remnant BHs \citep{Smith23LensedGW,AndreoniMargutti2024LSSTToO}. This can unlock unique scientific opportunities including the first glimpse of the GW source population at redshifts of $z\simeq1-2$ (Fig.~\ref{Fig:lensedGW}), unprecedented tests of General Relativity (GR) enabled by a step change in the precision of GW polarization measurements (e.g., \citealt{Goyal2021}), and a new ultra-precise probe of the Hubble tension (e.g., \citealt{Birrer2025}). Among the most exciting prospects is the very early detection of the lensed KN counterpart to the lensed GW when its \emph{second} image arrives at Earth with potential to be the earliest possible detection of a KN counterpart, and thus yield unprecedented constraints on the underlying physics \citep{Nicholl2024lessons}. 
\end{itemize}

\subsection{Follow-up of neutrino sources}
\label{SubSec:neutrinos}
LS4 will search for EM counterparts of neutrino sources.
For decades, there were only two examples of source-specific multi-messenger detections, both in MeV neutrinos: from the solar interior (starting in the 1960s), and from the nearby CCSN\,1987A~\citep{1987PhRvL..58.1490H}. The era of high-energy neutrino astronomy started with the discovery of an astrophysical diffuse flux of TeV-PeV neutrinos~\citep{2013Sci...342E...1I}. Roughly 10\% of that flux could be attributed to the Galactic plane~\citep{IceCube:2023ame}. The origin of the remaining, extragalactic flux is still an open question. Identified candidate neutrino sources include the close-by Seyfert galaxy NGC 1068~\citep{IceCube:2022der} and the flaring gamma-ray blazar TXS 0506+056 \citep{BlazarNeutrino}, but these individual sources can contribute at most a few percent to the diffuse flux. Furthermore, several extreme accretion flares (e.g., AT\,2019dsg, AT\,2019fdr and AT\,2019aalc;  \citealt{TDEneutrino,Reusch22TDEneutrino,Veres24TDEneutrino}) have been found in spatial and temporal coincidence with high-energy neutrino events. These accretion flares might be TDEs. While the neutrino emission from Seyfert galaxies is expected to be steady over time, the emission from blazars and accretion flares is expected to be variable or transient and is accompanied by a variable or transient EM counterpart that could be detected with LS4. Other neutrino source candidates with a distinct optical counterpart include interacting SNe \citep[e.g.,][]{2011PhRvD..84d3003M,2016APh....78...28Z,2017MNRAS.470.1881P,2018PhRvD..97h1301M,2019ApJ...874...80M,Wang:2019bcs,Sarmah:2022vra,Waxman:2024njn,Margutti09ip,Fang20}, choked-jet SNe~\citep[e.g.,][]{2001PhRvL..87q1102M,2005PhRvL..94j9903R,2008PhRvD..77f3007H,2013PhRvL.111l1102M,Nakar:2015tma,2016PhRvD..93h3003S}, and GRBs~\citep[see][and references therein]{Kimura:2022zyg}.

Different strategies exist to use neutrinos as triggers for electromagnetic follow-up observations. The challenge is to suppress the large atmospheric background that is present in water and ice Cherenkov detectors such as IceCube~\citep{IceCube:2016zyt} and KM3NeT~\citep{KM3Net:2016zxf}. IceCube publically releases a stream of single high-energy events (typically around 100\,TeV) with a high probability ($\gtrsim30\%$) to be of cosmic origin \citep{IceCube:2016cqr}. A recent update to this stream\footnote{\url{https://roc.icecube.wisc.edu/public/docs/IceCube_Update_Muon_Alert_Reco.pdf}} reduced the median 90\% angular uncertainty to $1.26$\,deg. 
Alternatively, spatial and temporal clusters of lower-energy neutrinos (1--10\,TeV) can be used to suppress the atmospheric background, which is mostly isotropically distributed in the sky. Short cluster (100--1000\,s) are selected to target short transients such as GRBs or choked-jet SNe \citep{IceCube:2018omy}, while longer time windows are used to search for TDEs, AGN flares or interacting SNe \citep{IceCube:2025qjb,IceCube:2016xci}. 

Current rates of neutrino alerts from IceCube are roughly 30\,yr$^{-1}$ for single high-energy events. Multiplets with maximal duration of 1000\,s, 30\,d and 180\,d will be released with an expected rate of a $\sim$few per year in the near future. Furthermore, detectors under construction are expected to contribute a realtime neutrino stream of $\mathcal{O}$(10\,yr$^{-1}$) during the run time of LS4. The scientific potential of partial detectors was illustrated by the detection of a 200\,PeV neutrino event by KM3NeT-ARCA with only $\sim10$\% of the detector deployed \citep{KM3NeTPeV}.

Ongoing observations and improvements in TeV-PeV neutrino observatories are likely to yield a large sample of astrophysical neutrinos, potentially enabling the detection of EM counterparts on a regular basis. IceCube is currently in operation, while Baikal-GVD and KM3NeT-ARCA are operating with partial detectors until they complete their arrays in the late 2020s. New neutrino telescopes are planned, which will expand the sky coverage, sensitivity, and energy range compared to current instruments. It is now imperative to expand the sample of neutrino sources with EM counterparts to confirm their physical associations with high statistical confidence, advance our understanding of these sources, and the physics of the neutrino emission.

\subsubsection{Comparison to existing/planned facilities in the Southern Hemisphere}
\label{SubSubSec:MMAsynergy}
To the best of our knowledge the future of DECam in 2025+ is uncertain. LS4 will work in synergy with Rubin-LSST, with the advantage of very flexible scheduling and larger field-of-view (but smaller collecting area). Two other time-domain surveys have a footprint in the southern skies: BlackGEM and ATLAS.
Differently from BlackGEM and ATLAS, the LS4 GW follow-up data will have a short proprietary period of $\sim$30\,d, allowing the entire astronomical community to benefit from our efforts. Following standard practice, our potential EM counterparts will be promptly announced through ATels, GCNs and/or AstroNotes, to enable time-critical follow-up observations on other facilities.


\section{Cosmology}

SNe\,Ia have famously been used as cosmological distance indicators, and they played an essential role in the discovery of the accelerating expansion of our Universe \citep{Riess98,Perlmutter99}.
SNe\,Ia remain a key cosmological probe, especially for measuring the recent expansion history of the Universe ($z\lesssim 1$), where dark energy drives the accelerating expansion \citep[for a review, see][]{2011ARNPS..61..251G}. Consequently, SN\,Ia data continue to provide strong constraints on dark energy \citep{Betoule14, Scolnic18, Pantheon, Union3, DES5yr}. In particular, current measurements aim to resolve the question of whether the dark energy density is constant (i.e., like
$\Lambda$), or a dynamical phenomenon involving new fields or modifications of the law of gravity beyond GR.

Recent results driven by SN\,Ia data \citep{Union3, DES5yr} combined with Cosmic Microwave Background \citep[CMB;][]{planck2020} and Baryon Acoustic Oscillation (BAO) constraints, show a 2.5--3.9\,$\sigma$ discrepancy with $\Lambda$ \citep[e.g.,][]{DESI1yr}. The dark energy equation of state can be parameterized as $w(a) = w_0 + w_a(1-a)$ (where $a = 1/(1+z)$ is the scale factor of the cosmic expansion), and these new measurements prefer models where $w_0 > -1$ and $w_a < 0$, implying that the dark energy density is dynamical, i.e., evolving with time. Adding to the challenge, there is a so-called ``Hubble tension'' where the local distance ladder measurements of $H_0$ based on using standardized stars (e.g., Cepheids, Tip of the Red Giant Branch, J-region Asymptotic Giant Branch) to calibrate distances to SN\,Ia host galaxies is 2--8\% larger than what is inferred from early Universe CMB observations, using $\Lambda$ plus cold dark matter ($\Lambda$CDM) and BAO measurements to extrapolate to the present time \citep{planck2020, Riess_2022, Freedman23, Freedman24}. This discrepancy has stubbornly resisted resolution. Dynamical dark energy or a break-down in the GR theory used for cosmological inference would point to new fundamental physics. Overall, deviations from the expectations of the $\Lambda$CDM model are strongest at low redshifts, making nearby SN\,Ia datasets such as that planned for the LS4 survey critical for fundamental physics.


\subsection{Anchoring the SN Hubble Diagram}
\label{sec:cosmo_anchor}

Continued study of this acceleration and the nature of dark energy will be conducted with large samples of SNe at high redshift by the forthcoming LSST (pushing to $z \sim 1$) and Roman (pushing to $z \sim 2$). To effectively measure cosmological parameters, these high-$z$ samples need to be anchored with hundreds or thousands of well-calibrated SNe\,Ia at low redshift. Recent efforts compiling such samples \citep[e.g.,][]{Union3,Rigault25}, are largely concentrated in the northern hemisphere and do not (yet) have the precision calibration that we plan to attain with LS4. A Hubble diagram composed of low-to-high redshift SNe discovered and observed by overlapping shallower and deeper surveys can be placed on a uniform magnitude system, mitigating a limiting source of uncertainty in current data. Even so, the relative motion of the local volume induced by large scale structure can imprint an offset on local SN brightnesses relative to those covering large volumes where this effect is averaged-out \citep{2006PhRvD..73l3526H}. An equally large ($N\sim4,000$) sample in the southern hemisphere will help to average-out this offset, which is one of the primary cosmology goals of LS4.

\subsection{Peculiar Velocities}
\label{sec:cosmo_pv}

The combination of SNe from the northern and southern hemispheres will also improve our ability to measure the power spectrum of galaxy/dark matter motions across the entire sky, which will, in turn, better constrain measurements of the growth of structure. The scatter along the redshift axis of the SN Hubble diagram induced by peculiar velocities is especially pronounced at lower redshifts. Spatial correlations in the velocity scatter are seeded by the same
dark matter field that gives rise to galaxy-count over-densities.  The peculiar velocity power spectrum is related to the dark matter power spectrum by a proportionality factor $P_{vv}\propto (fD)^2$, where $D$ is the spatially-independent ``growth factor'' in the linear evolution of density perturbations and $f \equiv \frac{d\ln{D}}{d\ln{a}}$ is the linear growth rate where $a$ is the scale factor  \citep{2006PhRvD..73l3526H,2011ApJ...741...67D}. When considering the local Universe it is common to express the growth factor as a constant normalized by a fiducial value, $D =\sigma_8/\sigma_{8,\text{fid}}$.  More generally, GR, $f(R)$ gravity \citep[e.g.,][]{Buchdahl70}, and Dvali-Gabadadze-Porrati (DGP) gravity \citep{Dvali00} follow the relation $f \approx \Omega_M^\gamma$ with $\gamma=0.55$, 0.42, and 0.68, respectively \citep{2007APh....28..481L}.  Using this parameterization to model gravity, peculiar velocity surveys probe $\gamma$ through $fD$, whose $\gamma$-dependence is plotted in Figure~2 of  \cite{1475-7516-2013-04-031}. SNe\,Ia with $z<0.1$ to first order measure the strength of gravity and with redshift-dependence can distinguish between GR and competing theories of modified gravity.

Over its full duration, LS4 will discover and measure light curves of $\sim$4,000 SNe\,Ia out to $z \approx 0.1$. The relatively sparse cadence and large aperture for LSST WFD are not well suited for precision light curves of low-$z$ SNe, as shown in Figure~\ref{fig:synergy}. LS4 will do considerable work to fill in the gaps in low-$z$ LSST observations with both the LEG and FOOT surveys, which will enable (i) robust identifications of SN\,Ia candidates well before peak in order to obtain a spectroscopic classification when the transient is brightest, and (ii) easy photometric classifications via the $i$ and $z$ band secondary maxima, prominent in normal SNe\,Ia, for those without spectra. Thus, LS4 SNe\,Ia will tightly constrain the growth of structure in the low-redshift Universe. 

Projections using the {\textit{FLIP}} code of \cite{2025arXiv250116852R} indicate that 4,000 SNe\,Ia to $z \approx 0.1$ over the LS4 footprint can produce 13/15/22\% uncertainties in $f\sigma_8$ from peculiar velocities alone and 9/11/15\%  uncertainties when combined with the galaxy density field from redshift surveys, respectively for $0.08$/$0.11$/$0.15$\,mag standardization dispersion. The dispersion realized for LS4 will depend in part on the extent and quality of spectroscopic follow-up we are able to obtain, or the standardization improvements our redder bandpasses may offer. Studies indicate that the photometric misclassification of SNe\,Ia with LSST-quality light curves will induce a bias in $f\sigma_8$ measurements that is less than the statistical error for $z<0.16$ for the case of 0.15\,mag standardization dispersion \citep{rosselli}. Thus, we can expect that a survey relying solely on joint LS4$+$LSST discoveries and photometry coupled with spectroscopic host-galaxy redshifts would be competitive, but also has the potential to be improved substantially. Our forecast values are comparable to the 15.3/6.5\% velocity-only and velocity$+$galaxy results anticipated for DESI \citep{Saulder23, Said24} and the 5\% velocity+galaxy forecasts for 4MOST \citep{4MOSTHS23}, calculated using (somewhat more optimistic) Fisher matrix forecasts for the Fundamental Plane and Tully-Fisher galaxy-based distance indicators.

We note that these efforts are complementary to the ongoing ZTF survey, since ZTF and LS4 are in different hemispheres covering different parts of the sky. Combining SNe from the two surveys roughly doubles the survey volume and increases by 4 the number of SN pairs, tightening the $f\sigma_8$ uncertainty by a factor $\sqrt{2}$.

LS4 will provide a low-redshift anchor on $f\sigma_8$ at $0<z<0.1$ whereas precise measurements at higher redshifts will come from redshift space distortions derived from galaxy surveys \citep{2017ApJ...847..128H}.
When both probes are combined, the redshift-dependent evolution
of $f\sigma_8$ will provide a strong test
of GR and other gravitational models
\citep{PhysRevD.101.023516}.

\subsection{Reducing the Uncertainty on SN\,Ia Distance Measurements}
\label{sec:cosmo_systematics}

Nearby SNe\,Ia are bright enough that their diversity can be studied in detail, enabling the investigation of potential astrophysical biases that could affect the inference of cosmological parameters. For such nearby SNe\,Ia, LS4 will provide $z$-band light curve observations that are currently rare, and more $i$-band observations (relative to ZTF). These data can help solve the complex origin of SN--host correlations by providing a longer wavelength baseline, e.g., for studying the effects of host-galaxy dust, while also improving photometric classification by exploiting the $i$ and $z$ band secondary maxima seen in SN\,Ia light curves \cite[e.g.,][]{Riess96, Kasen06}. Furthermore, the overlap in time and sky coverage with LSST enables extended phase coverage (see e.g., Figure~\ref{fig:synergy}). These synergistic observations will provide an unprecedented opportunity to study complex nonlinear and color-dependent standardization effects.
 
SN\,Ia are empirically calibrated distance indicators because their progenitors and explosion mechanism are insufficiently known, and predictive models at the required few-mmag level do not yet exist. Light-curve standardization reduces the natural dispersion of $\sim$0.4\,mag to 0.12---0.15\,mag \citep{Phillips93, Tripp98, Riess96}. Spectroscopic standardization \citep{BooneTE2021ApJ...912...71B, Stein_2022ApJ...935....5S} and some NIR data \citep{2012MNRAS.425.1007B} can reduce this scatter to $\sim$0.08\,mag. The remaining scatter is unexplained by measurement or modeling uncertainties, leaving ample room for unaccounted effects that could bias the inferred distances. Thus, the uniformity and differences in the observed calibration parameters of SNe\,Ia as a function of redshift and host environment are an area of active investigation in order to ensure they are sufficiently reliable standardizable candles for measuring dark energy and $H_0$.

Within this context, large samples have clearly demonstrated that the SN\,Ia standardized brightness correlates with host properties. For example, SNe\,Ia in massive galaxies are $\sim$0.1\,mag brighter than those in low-mass galaxies. \citep{Kelly10, Sullivan10, Childress13, Rigault2013}. Further investigation of this correlation has suggested it could originate from a progenitor effect, such that SN\,Ia associated with young stellar environments are intrinsically fainter \citep{Rigault2013,Rigault20}, or variations in host galaxy dust properties along the SN\,Ia line of sight \citep{Brout21, Johansson_2021}, or both. The offset in magnitude between SNe\,Ia in low and high-mass hosts (often referred to as the ``mass-step'') has been used as a third standardization parameter in recent cosmological analyses \citep[e.g.,][]{Betoule14,Scolnic18,Riess_2022,Union3,DES5yr}. This correction is applied despite the ``mass-step'' likely being a proxy for a more fundamental effect driving the magnitude offset. Spectroscopic standardization reduces such ``steps'' by about 2--3$\times$ \citep{BooneTE2021ApJ...912...71B}, indicating that much of the effect arises from the limitations of the present-day light-curve standardization method itself, and further indicating that  SNe\,Ia do in fact transmit the information necessary to improve their standardization. 

Such studies, with the most recent examples from the large ZTF DR2 dataset include: confirming that the stretch parameter commonly used to account for the luminosity-decline rate relation \citep{Phillips93, Perlmutter97} is non-linear \citep{Amanullah10, Rubin15, Ginolin25}; clearer evidence of the expected intrinsic SNe\,Ia color component in addition to reddening by dust \citep{Mandel17, Mandel22, Union3, Ginolin25}; even larger ``steps'' than the canonical ``mass-step'' using different indicators and/or unbiased samples \citep{Rigault20, Ginolin25}. Even for the host-galaxy dust component, cosmology applications must consider $R_V$ variations \citep{Amanullah+15, Mandel17, Huang17, Mandel22, Brout21, Union3, grayling24, Ginolin25}.

Finally, we note that other contemporaneous surveys will enrich the LS4 SN cosmology dataset. LSST over the full LS4 footprint, and Euclid and Roman over some of the footprint, will provide exquisite, deep, higher resolution imaging, in more bands, of the LS4 host galaxies. 4MOST and DESI will provide high-quality spectra of many of these same host galaxies. Combined, these offer a significant advantage over many previous studies of SN\,Ia standardization residuals correlated with global and local host galaxy properties.

\subsection{Cosmology with SNe\,II-P}
\label{sec:cosmo_snIIp}

SNe\,II as cosmological probes have lagged behind their brighter and better calibrated cousins, SNe\,Ia, but they nevertheless hold cosmological value. Recent studies have used new samples of SNe\,II and demonstrated that the subset with long photometric plateaus, SNe\,II-P, have outstanding cosmological utility (for a review, see \citealt{2017hsn..book.2671N}).

There are multiple advantages for using SNe\,II to probe cosmology over SNe\,Ia: (i) the progenitor systems of SNe\,II-P are well understood; (ii) SNe\,II-P are more numerous per unit volume by a factor of 3; and (iii) the plateau phase lasts for $\sim 100$ days, allowing ample time to take a spectrum of the event and measure its expansion velocity. Recently, the so-called Standard Candle Method \citep[SCM;][]{Hamuy02, Nugent06, Olivares10} achieved $\sigma_M=0.27$~mag with nearly 90 nearby SNe\,II-P \citep{2020MNRAS.496.3402D}. Given the advantage in number density, the relative weight of SNe\,II-P is comparable to SNe\,Ia for the peculiar velocity measurements ($\sigma_M=0.12$--0.15\,mag). These measurements depend on constraining on the explosion date. In the traditional expanding photosphere method \citep[e.g.,][]{leonard02-99em}, one can overcome this hurdle by taking multiple spectra \citep{KK74} and simultaneously solving for the luminosity and explosion date. However, with the LS4 FOOT cadence, we will be able to constrain the explosion dates to within a day, obviating this requirement and allowing facilities like DESI \citep{DESI-Collaboration16} and 4MOST/TiDES to leisurely take a single spectrum during the plateau phase. When coupled with the exquisite multi-color photometry from LS4, distance measurements out to $z=0.05$ will be straightforward, not only improving peculiar velocity measurements, but also contributing to an independent measurement of $H_0$ (\citealt{2022MNRAS.514.4620D}; see \citealt{2023arXiv230517243D} for a review).

Similarly as for the empirical standardization method of SN\,Ia, the SN\,II-P empirical standardization method will need to be examined for residual correlations with host-galaxy properties. Alternatively, a more ``first-principles'' method such as the expanding photosphere method \citep{KK74, Schmidt94, Vogl24} could be used. 

\subsection{Precise Instrumental Throughput Calibration}
\label{sec:cosmo_cal}

The precise determination of luminosity distances to objects at different redshifts requires knowing the instrumental response function of the astronomical instrument, and making correspondingly precise corrections for extinction along the line of sight to the source. Reaping the benefit of improved statistics from large increases in the number of detected SNe requires lowering the floor on systematic errors due to photometric calibration \citep{stubbs2015precise}. 

The sensitivity function $S(\lambda)$ of an instrument establishes both the zeropoints and the bandpass shapes. Zeropoints are typically determined using standard stars, such as the CALSPEC suite of spectrophotometric standards \citep{bohlin2020new}, and the more recently established network of fainter white dwarf standards \citep{boyd2024damodel}. However, these do not provide the bandpass shapes with sufficient fidelity for cosmology with SNe. Moreover, cosmology with SNe relies on relative distances --- so it is the filter-to-filter zeropoints that need to be determined at the few--mmag level.

A method to determine both the relative filter zeropoints and the bandpass shapes exploits the fact that the quantum efficiency of silicon photodiodes can be determined at the part-per-thousand level. Using this as the metrology basis for photometry requires transferring that calibration knowledge onto the astronomical imaging system \citep{stubbs2006toward}. This has now been shown at the few-mmag level \citep{souverin2024stardice}. 

The LS4 survey will utilize both celestial and terrestrial calibrators to establish the relative zeropoints across the ${\it giz}$ passbands. We will deploy PANDORA, a device that injects a known dose of monochromatic photons into the telescope pupil, to measure the instrumental response function \citep{sjoberg2025pandora}. This knowledge will be supplemented with a comparison of synthetic and on-sky photometry of white dwarf calibration objects, and all-sky cross-calibration with the Rubin photometric system. On occasion there will be LS4 SN observations simultaneous with LSST, greatly strengthening the calibration between the SN samples from these two surveys.


\section{Stellar Lensing, Flares and Variability}

Select fields in the Milky Way Galactic bulge and in the Magellanic Clouds have been regularly monitored for decades with a comparable depth and cadence to what is planned for LS4 \citep[e.g., OGLE, the Massive Compact Halo Objects project; MACHO, EROS, the Korea Microlensing Telescope Network; KMTNet, the Microlensing Observations in Astrophysics survey; MOA;][]{Alcock96a,Udalski92,Ansari96, kmtnet:kim:2018, moa:Sumi:2003}. The \textit{entire} southern sky has also been observed repeatedly by surveys such as Gaia, Transiting Exoplanet Survey Satellite (TESS), ASAS-SN, and ATLAS.  The cadence from Gaia is sparse compared to LS4 and the bandpasses are fewer and non-standard. Surveys such as ASAS-SN and TESS observe with higher cadence, but, critically, the LS4 survey will reach single-epoch depths $\sim$5~mag deeper than these precursors. As such, the same Galactic classes found in ASAS-SN or TESS could be observed at ten times the distance within a thousand times larger volume. In its first three years of operation ZTF found $\mathcal{O}(10^6)$ variables \citep{2020arXiv200701537O}, and we expect to find and characterize a similar number with LS4.

\subsection{Microlensing}
Gravitational microlensing occurs when a massive ``lens'' object passes near the line of sight between a luminous ``source'' object and an observer and the lens bends the light path from the source. This produces a transient brightening (\textit{photometric} microlensing) and apparent positional shift (\textit{astrometric} microlensing) of the source star. Characterizing these events enables direct measurement of the mass of the lens. Thus, microlensing is an ideal method for detecting compact objects, dim stars, brown dwarfs, and cool planets. Within the LS4 SOLE survey of the Galactic plane and bulge, photometric microlensing events will be discovered. Particularly interesting events (e.g., BH candidates) may be followed up with high-resolution imaging from space or with ground-based adaptive optics to measure the astrometric signal during the event, or as late-time follow-up to observe the lens-source separation \citep[e.g.][]{Lu2016, Abdurrahman2021}. 

In gravitational lensing events, the source's light is bent around the lens' Einstein ring, and parametrized by the angular size $\theta_E$, the Einstein ring radius crossing time $t_E$, and the microlensing parallax $\pi_E$. Observation of photometric microlensing measures $t_E$ and $\pi_E$, but additional information is typically required to break the mass-distance degeneracy and fully characterize the lens system. For the study of stellar mass black holes, we can use follow-up astrometry to measure $\theta_E$ and determine the lens mass. One isolated black hole has been detected with this technique so far \citep[OGLE-2011-BLG-0462,][]{Lam2023, Sahu2022}, and LS4 will help identify new candidates to build a sample of several black holes.

While LSST will be capable of finding microlensing events, the WFD cadence is not ideal for characterizing average microlensing events \citep[see][]{Abrams2025}. A $\sim$1~d cadence is critical to fully measure the shape of the microlensing light curve and the microlensing parallax \citep[see, e.g.][]{LamThesis}. For events detected in both surveys, LSST will provide photometry with higher precision and additional filters, improving the output of multi-band event fitting.

\subsubsection{In the Galactic plane}
\begin{figure}
    \centering
    \includegraphics[width=3.2in]{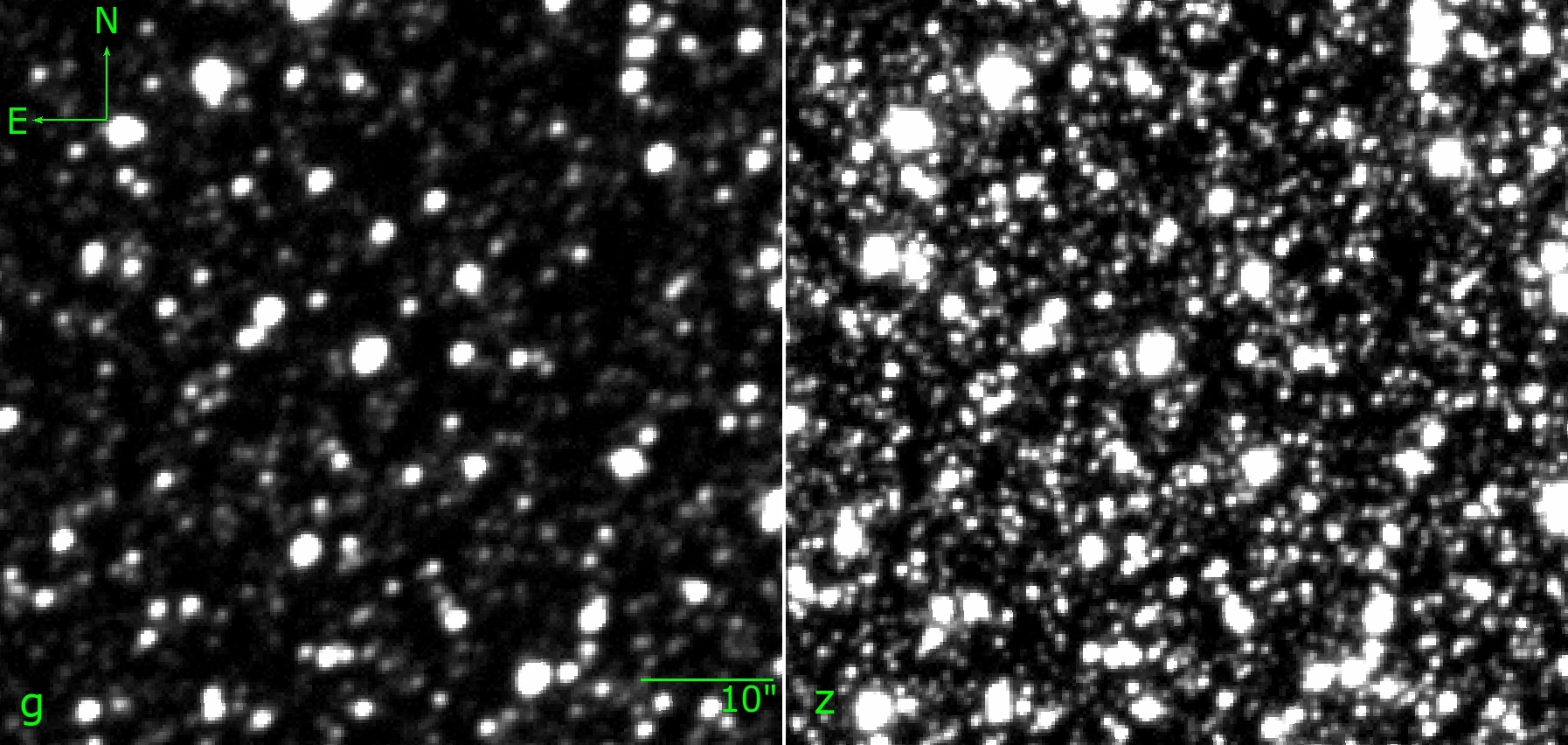}
    \caption{Comparison of the number of stars that can be observed in blue vs.\ red filters for a high-extinction field toward the Galactic plane. Images are $60'' \times 60''$ cutouts from a DECaPS-East deep drilling field \citep{Graham2023} located at right ascension and declination $\alpha_\mathrm{J2000} = 07^\mathrm{h} \, 45' \, 16\farcs8, \, \delta_\mathrm{J2000} = -26^\circ \, 15' \, 00\farcs0$. North is up and east is to the left.     \textit{Left}: a 96\,s $g$-band exposure with DECam. \textit{Right}: a 30\,s $z$-band exposure of the same field that reveals a great deal more stars than the longer exposure $g$-band observation. 
    }
    \label{fig:galactic_gz}
\end{figure}

The characteristics of microlensing events are determined by the spatial and kinematic distributions of luminous objects and massive objects in the Milky Way, which probes the structure of the Galaxy. To date, the majority of ongoing microlensing surveys have primarily been focused on the Galactic bulge and the areas around it. The LS4 SOLE survey offers a valuable opportunity to search for microlensing throughout the plane from a southern observing site. Furthermore, $z$-band observations will allow LS4 to peer deeper into the most extinguished regions in the Galactic plane than surveys that only extend redward to the $r$ and/or $i$ bands (see Figure \ref{fig:galactic_gz}). LS4's multi-filter configuration will provide useful color information since microlensing events are achromatic to first order, unlike many other forms of variability.

Several studies have analyzed microlensing events detected in ZTF observations in and out of the Galactic plane \citep[e.g.,][]{Mroz2020,Rodriguez2022,Medford2023,Zhai2025}, noting the dependence of microlensing event timescales on Galactic location. This work has highlighted how surveys of the plane can help characterize dark objects residing in the Galactic disk. 

By mapping the microlensing event rates and characteristics throughout the Galactic bulge and plane, we can map the density of free-floating BHs as a function of Galactic longitude and latitude \citep[see e.g.][]{Wyrzykowski2016,Lam2020,Perkins2024}, determine whether BHs and NSs receive kicks at formation \citep[see e.g.][]{Rose2022,Koshimoto2024}, compare the mass functions and binary fractions of stars in different environments \citep[see e.g.][]{Wegg2017, Chabrier2023, Abrams2025a}, and determine whether planet occurrence depends on Galactic environment \citep[see e.g.][]{Penny2016, Koshimoto2021}. While the planned 1~d cadence for the LS4 SOLE survey is not fast enough to fully characterize exoplanet events, targeted higher-cadence follow-up will enable measurements of the planetary properties \citep[see e.g.,][]{Bachelet2024}.

\subsubsection{Off the Galactic plane}

The targeted discovery of microlensing events toward the bulge, M31, and Magellanic Clouds has been critical for constraining dark matter candidates and understanding the distribution of planets at large distances from their host star. The LS4 LEG and FOOT surveys, while generally focused away from regions of large optical depth to microlensing, is highly favorable for the characterization of individual microlensing events. Departures on the order of a few days from single-lens curves due to planetary lensing will be readily detectable. \citet{Medford2023} identified 19 microlensing event candidates at $|b|\geq10^\circ$ from the first three years of ZTF, almost doubling the number of events discovered in the stellar halo. LS4 should similarly discover microlensing events away from the plane. The Einstein crossing time statistics of events off the plane can potentially constrain the contribution of primordial BHs to the dark matter mass fraction \citep{2016PhRvD..94f3530G}.

\subsection{Other Galactic Science}

Long-duration, high-cadence observations, such as those collected by CRTS and ZTF, have discovered many short-period ($<90$,min) systems \citep{2014ApJ...790..157D,2020arXiv200701537O}, including eclipsing dwarf novae, AM CVn stars \citep{2018MNRAS.475.2560V}, double degenerate systems \citep{2020MNRAS.494L..91C} and below-the-period gap cataclysmic variables (CV). We expect LS4 to uncover a bevy of such systems, including LISA verification binaries---those with well-constrained GW strain inferences \citep[e.g.,][]{2018MNRAS.480..302K}. In addition to microlensing and GW follow-up, LS4 will enhance BH science by detecting optical counterparts to X-ray binaries \citep[see e.g.][]{Malzac2003}. BHs and other compact objects with detached luminous companions can be detected via periodic optical signals from self-lensing, tidal ellipsoidal variation, and relativistic beaming \citep[see e.g.][]{Chawla2024, Nir2025}.

As found in repeated Stripe 82 observations to $V \sim 21$ \citep[][similar to LS4]{2007AJ....134.2236S}, RR\,Lyrae can probe the inhomogeneous mass distribution \citep{2001ApJ...554L..33V} of the outer halo of the Milky Way to $\sim$120\,kpc, revealing (due to the higher population density in the inner halo) $\sim$1.5\,kpc substructure  \citep[such as tidal streams;][]{2003ApJ...596L.191N} out to 30\,kpc. Using RR\,Lyrae as probes, our understanding of the kinematics of the outer halo, inner halo and disk can be constrained by LS4 \citep[e.g.,][]{2018ApJ...859...31H,Iorio21,Medina24}.

\section{Summary}

In this paper we have described the design of the LS4 LEG, FOOT, and SOLE surveys, which will monitor the southern sky at cadences ranging from 1 to 3\,d. The LS4 LEG survey will tile the extragalactic sky with a 3\,d cadence to discover SNe\,Ia to be used in cosmological analyses, TDEs, AGN, and rare explosive transients. Nightly monitoring by the LS4 FOOT survey will reveal explosive transients shortly after they explode. The LS4 SOLE survey will map Galactic fields, primarily toward the bulge and inner plane, to find microlensing events and short-period binary systems. The LS4 partnership will also conduct ToO observations of GW events discovered by LVK and monitor the Euclid and ULTRASAT deep fields with the 10\% of the observing time that is not part of the public survey. 

The landscape of time-domain astronomy is rapidly evolving. Whereas one to two decades ago it was possible for individual collaborations to have a major impact with a single wide-field, optical telescope, modern efforts to identify transients and variables must do so in an environment where several other telescopes are likely to be searching the same footprint on the sky. Later this year, any search or monitoring in the southern hemisphere will overlap LSST WFD observations, meaning smaller aperture surveys must compete and/or complement the deep, but at times sparse, observations from LSST. Furthermore, several wide-field transient surveys to be conducted outside the optical are planned or already underway, with projects like Einstein Probe \citep{Yuan18} monitoring the high-energy sky, the Karl~G.~Jansky Very Large Array Sky Survey \citep[VLASS;][]{Lacy20} searching for radio transients, ULTRASAT and UVEX searching the UV, and Roman finding high-redshift transients with deep infrared imaging. Taken together, we contend that future progress in the field will require clever strategies to complement the observations of other facilities and combine alerts from different surveys that span the EM spectrum. LS4 is an attempt to design such an experiment: while we have outlined a lot of science that LS4 can accomplish on its own, many of the most exciting science goals are directly the result of combining LS4 with substantially deeper surveys such as Rubin, Euclid, Roman, and ULTRASAT. Leveraging these disparate efforts affords LS4 the opportunity to have an outsized impact despite its modest (aperture) size. 

\vspace{5mm}
\textit{Acknowledgements} --- \\
{
This research used resources of the National Energy Research Scientific Computing Center (NERSC), a Department of Energy Office of Science User Facility using NERSC award HEP--ERCAP33561.  PEN, RAK, KWL, and CW acknowledge support from the DOE/ASCR through DE-FOA-0001088, Analytical Modeling for Extreme-Scale Computing Environments, the X-SWAP Project.

AAM, CL, and NR are supported by DoE award no.~DE-SC0025599.
IA acknowledges support from the European Research Council (ERC) under the European Union’s Horizon 2020 research and innovation program (grant agreement no.~852097), from the Israel Science Foundation (grant no.~2752/19), from the United States - Israel Binational Science Foundation (BSF; grant no.~2018166), and from the Pazy foundation (grant no.~216312).
NSA acknowledges support from the National Science Foundation under grant no.~1909641 and the Heising-Simons Foundation under grant no.~2022-3542.
SA is supported by an LSST-DA Catalyst Fellowship (Grant 62192 from the John Templeton Foundation to LSST-DA). SA also gratefully acknowledges support from Stanford University, the United States Department of Energy, and a generous grant from Fred Kavli and The Kavli Foundation.
DB is partially supported by a NASA Future Investigators in NASA Earth and Space Science and Technology (FINESST) award no.~80NSSC23K1440.
FEB acknowledges support from ANID-Chile BASAL CATA FB210003, FONDECYT Regular 1241005, and Millennium Science Initiative, AIM23-0001.
HB acknowledges the support by ANID BASAL Project 210003, ANID grant: Programa de Becas/ Doctorado Nacional (21241862).
Support for MC is provided by ANID's FONDECYT Regular grant no.~1171273; ANID's Millennium Science Initiative through grants ICN12\textunderscore 009 and AIM23-0001, awarded to the Millennium Institute of Astrophysics (MAS); and ANID's Basal project FB210003.
AF and PMV acknowledge the support from the DFG via the Collaborative Research Center SFB1491 \textit{Cosmic Interacting Matters - From Source to Signal}.
CF acknowledges support from STFC funding through grants ST/V002031/1, ST/X00130X/1, and the Royal Society through grant IES\textbackslash\,R3\textbackslash\,223075.
AG acknowledges support from  the Swedish Research Council, Dnr 2020-03444 and the Swedish National Space Agency, Dnr 2023-0022.
CPG acknowledges financial support from the Secretary of Universities and Research (Government of Catalonia) and by the Horizon 2020 Research and Innovation Programme of the European Union under the Marie Sk\l{}odowska-Curie and the Beatriu de Pin\'{o}s 2021 BP 00168 programme, from the Spanish Ministerio de Ciencia e Innovaci\'{o}n (MCIN) and the Agencia Estatal de Investigaci\'{o}n (AEI) 10.13039/501100011033 under the PID2023-151307NB-I00 SNNEXT project, from Centro Superior de Investigaciones Cient\'{i}ficas (CSIC) under the PIE project 20215AT016 and the program Unidad de Excelencia Mar\'{i}a de Maeztu CEX2020-001058-M, and from the Departament de Recerca i Universitats de la Generalitat de Catalunya through the 2021-SGR-01270 grant.
LG acknowledges financial support from the Spanish Ministerio de Ciencia e Innovaci\'{o}n (MCIN) and the Agencia Estatal de Investigaci\'{o}n (AEI) 10.13039/501100011033 under the PID2023-151307NB-I00 SNNEXT project, from Centro Superior de Investigaciones Cient\'ificas (CSIC) under projects PIE 20215AT016, ILINK23001, COOPB2304, and the program Unidad de Excelencia Mar\'ia de Maeztu CEX2020-001058-M, and from the Departament de Recerca i Universitats de la Generalitat de Catalunya through the 2021-SGR-01270 grant.
IH gratefully acknowledges support from the Leverhulme Trust [International Fellowship IF-2023-027] and the UKRI Science and Technologies Facilities Council [grants ST/V000713/1 and ST/Y001230/1].
MJH acknowledges support from the Heising-Simons Foundation under grant no.~2022-3542.
SH was supported through a NASA grant awarded to the Illinois/NASA Space Grant Consortium.
CDK gratefully acknowledges support from the NSF through AST-2432037, the HST Guest Observer Program through HST-SNAP-17070 and HST-GO-17706, and from JWST Archival Research through JWST-AR-6241 and JWST-AR-5441.
LAK is supported by a CIERA Postdoctoral Fellowship.
RL acknowledges funding by the European Union (ERC, project no.~101042299, TransPIre). Views and opinions expressed are however those of the author(s) only and do not necessarily reflect those of the European Union or the European Research Council Executive Agency. Neither the European Union nor the granting authority can be held responsible for them.
JRL acknowledges support from 
Heising-Simons Foundation Award no.~2022-3542.
DM acknowledges support by ERC Grant no.~833031 from the European Union. 
KM acknowledges funding from Horizon Europe ERC grant no.~101125877.
LM acknowledges support through a UK Research and Innovation Future Leaders Fellowship (grant no.~MR/T044136/1).
RM acknowledges support by the National Science Foundation under award no.~AST-2224255.
MN is supported by the European Research Council (ERC) under the European Union’s Horizon 2020 research and innovation programme (grant agreement no.~948381).
GP acknowledges support from the National Agency for Research and Development (ANID) through the Millennium Science Initiative Program – ICN12\_009.
JLP acknowledges support from ANID, Millennium Science Initiative, AIM23-0001.
BDS acknowledges support through a UK Research and Innovation (UKRI) Future Leaders Fellowship [grant no.~MR/T044136/1].
CWS thanks Harvard University for support of this effort.
GPS acknowledges support from The Royal Society, the Leverhulme Trust, and the Science and Technology Facilities Council (grant no.~ST/X001296/1).
MTS thanks the Israeli Science Foundation (grant nos. 2068/22 and 2751/22).
NS\ acknowledges support from the Knut and Alice Wallenberg Foundation  through the ``Gravity Meets Light'' project and by and by the research environment grant ``Gravitational Radiation and Electromagnetic Astrophysical Transients'' (GREAT) funded by the Swedish Research Council (VR)  under Dnr 2016-06012.
BT acknowledges support from the European Research Council (ERC) under the European Union's Horizon 2020 research and innovation program (grant agreement no.~950533) and from the Israel Science Foundation (grant no.~1849/19).
CW acknowledges support from the LSST Discovery Alliance under grant AWD1008640. 

Fermi National Accelerator Laboratory (Fermilab) is a US Department of Energy, Office of Science, Office of High Energy Physics HEP User Facility. Fermilab is managed by FermiForward Discovery Group, LLC, acting under Contract no.~89243024CSC000002.
This work was supported by the US Department of Energy (DOE), Office of Science, Office of High-Energy Physics, under contract no.~DE–AC02–05CH11231.
We gratefully acknowledge funding from ANID grants: Millennium Science Initiative - AIM23-0001 (FEB); CATA-BASAL - FB210003 (FEB); and FONDECYT Regular - 1241005 (FEB).
This research was funded in part by the Koret Foundation and by the Kavli Institute for Particle Astrophysics and Cosmology at Stanford University.
This work was supported by ANID, Millennium Science Initiative, AIM23-0001, and Centro de Modelamiento Matemático (CMM) BASAL fund FB210005.

The material contained in this document is based upon work partially supported by a National Aeronautics and Space Administration (NASA) grant or cooperative agreement. Any opinions, findings, conclusions, or recommendations expressed in this material are those of the author and do not necessarily reflect the views of NASA.

Major funding for LS4 has been provided by the LS4 Founding Members: Northwestern University, Bar-Ilan University, Fermi National Accelerator Laboratory, Lawrence Berkeley National Laboratory, the Millennium Institute of Astrophysics, University of Portsmouth Higher Education Corporation, Tel Aviv University, The Regents of the University of California, Berkeley, and Yale University. Project operations are covered by: DESY, IN2P3, Lancaster University, Purdue University, Ruhr-Universit\"{a}t – Bochum, Stockholm University, Trinity College Dublin, University of Southampton, Institute of Space Sciences (ICE, CSIC), Queen’s University Belfast, and the University of Birmingham.
}

\vspace{5mm}
\facilities{ESO:Schmidt}


\software{\texttt{matplotlib} \citep{Hunter07}, 
\texttt{pandas} \citep{pandas20},
\texttt{sncosmo} \citep{Barbary16}
          }
          

\newcommand{\noop}[1]{}




\end{document}